\documentclass[journal,12pt,onecolumn,draftclsnofoot,]{article}
\usepackage{geometry}
\usepackage{setspace}   
\usepackage{adjustbox}
\usepackage{hyperref}
\usepackage{xcolor}
\usepackage{siunitx}
\usepackage{xparse}%
\usepackage{cite}
\usepackage{authblk}
\usepackage{multicol}
\usepackage{booktabs}
\usepackage{subfig}
\usepackage{xcolor,colortbl}
\usepackage[T1]{fontenc}
\usepackage{lipsum}
\usepackage[utf8]{inputenc}
\usepackage{array}
\usepackage{lettrine}
\usepackage{booktabs}
\usepackage{authblk}
\usepackage{graphicx}
\usepackage{stfloats}
\usepackage{amsmath}
\usepackage{graphicx}
\usepackage{amssymb}
\usepackage[round, sort, numbers]{natbib}
\usepackage[mathletters]{ucs}
\usepackage[skip=2pt,font=footnotesize]{caption}

\begin{document}
	\newgeometry {top=25.4mm,left=19.1mm, right= 19.5mm,bottom =19.5mm}%

\title{
Fractional Order Models of Arterial Windkessel as an Alternative in the Analysis of the Left Ventricular Afterload}
\author{Mohamed~A. Bahloul$^1$,
        and Taous-Meriem Laleg Kirati$^{1,2}$ }
    \affil{$^1$ Computer, Electrical and Mathematical Sciences and Engineering Division, King Abdullah University of Science and Technology, Makkah Province, Saudi Arabia\\
    $^2$ French national research institute for the digital sciences, INRIA, Paris, France \\ E-mail:\{mohamad.bahloul, taousmeriem.laleg\}$@$ kaust.edu.sa}
\date{}
\maketitle

\doublespacing

\begin{abstract}
In this paper, a new fractional order generalization of the classical Windkessel arterial model is developed to describe the aortic input impedance as an assessment of the left ventricular after-load. The proposed models embeds fractional-order capacitor to describe the total arterial compliance.
In this paper, we report our investigations on fractional calculus tools and demonstrate that fractional-order impedance can be used to determine the vascular properties and studying its dynamic effects.
We conceived two fractional-order lumped parametric models: the fractional-order two-element Windkessel model and the fractional-order three-element Windkessel model. We compared these models to the classical Windkessel one using in-silico ascending aortic blood pressure and flow database of 3,325 virtual subjects. Results showed that the proposed fractional-order models overcame the limitations of the standard arterial Windkessel model and captured very well the real dynamic of the aortic input impedance modulus. We also demonstrated that the proposed models could monitor the changes in the aortic input impedance for various arterial physiological states. Therefore, our models provide a new tool for “hemodynamic inverse problem” solving and offer a new, innovative way to better understand the viscoelastic effect, in terms of resistive behavior of the arterial motions.
\end{abstract}


\section{Introduction}
The arterial system is entirely coupled with the heart, so that the contractile state of the left ventricle and its produced central blood pressure (the pressure in the aorta) are in tune with the arterial mechanical properties \cite{borlaug2011ventricular}. The interactions between the left ventricle and the systemic arteries are considered of imperative relevance in governing appropriate and normal cardiovascular function \cite{o1984physiological}. Over the last century, numerous methods have been proposed to characterize the complex vascular after-load presented by the systemic arteries to the left ventricle; known as the aortic input impedance, \cite{milnor1975arterial,sharp2000aortic}. In general, it is  very challenging to measure such vascular parameter directly. However, nowadays, blood pressure and flow waveforms at the arterial entrance (or alternatively aortic input impedance, which is expressed as the ratio of the blood pressure and flow in the frequency domain) are accessible practically. Accordingly, the derivation of the arterial mechanical properties from measured blood pressure and flow, at the entrance of the systemic system, is possible  \cite{westerhof2019arterial}. This is equivalent to solve a single equation with two known inputs: blood flow and pressure (equivalently the aortic input impedance) and multiple unknowns (mechanical parameters). A typical approach in solving this so called “hemodynamic inverse problem” is based on fitting the real input impedance to a reduced model and then the resulting estimated parameters represent the vascular properties \cite{quick2006resolving,quick2001infinite}. In the open literature review, several reduced models have been proposed in this regard \cite{shi2011review}. The well-known Windkessel $\mathrm {(WK)}$ lumped parametric model has been considered, for a long time, as an acceptable approximation of the aortic input impedance \cite{westerhof2019arterial}. WK models  reduced number of unknown parameters,  are able to fit the real input impedance at low and high frequencies and involve physiologically interpretable elements. However, the WK model presents some limitations, such as the inability to represent all the arterial mechanical properties of interest accurately such as arterial stiffness \cite{capoccia2015development}. 

Similarly to any bio-tissue, the systemic arteries present a viscoelastic behavior , not an entirely pure elastic one \cite{vcanic2006modeling,holzapfel2002structural,wang2016viscoelastic,balocco2010estimation,burattini1999viscoelasticity}. However, most of the proposed $\mathrm {WK}$ models regard the arteries as a pure elastic reservoir, modeling the total arterial compliance by an ideal capacitor with a value that is constant over the whole cardiac cycle. Some research attempts have been made to describe the arterial viscoelasticity properties by connecting a small resistance in series with the previous ideal capacitor, resulting in the creating of a complex and frequency-dependent compliance. The later configuration is based on the electrical analog of the Voigt mechanical cell that consists of a spring connected in parallel to a dashpot, accounting for the static compliance and viscous losses, respectively \cite{burattini1998complex,aboelkassem2019hybrid}. Even though, many studies have argued that the Voigt representation is a very poor configuration of the vascular viscoelasticity, since it does not account for stress-relaxation experiment. Yet,this representation is commonly used in Viscoelastic Windkessel models $\mathrm {(VWK)}$. This is related to the fact that, even though higher-order viscoelastic models would provide a more natural and realistic representation, real aortic input impedance cannot depict sufficient information to identify all the parameters of these complex models\cite{burattini1998complex}.

 Generally, conventional integer-order lumped parameter models used to simulate the viscoelastic properties of such bio-tissues are not sufficient, as they do not account for the power law demonstrated experimentally in viscoelastic material \cite{naghibolhosseini2018fractional}. The power-law like stress-relaxation, observed experimentally in viscoelastic materials, is expected to be seen in vascular tissue\cite{hemmer2009role}. Recently, a fractional-order derivative operator that can be defined as a generalization of the standard integer derivative to a non-integer order, has received considerable interest in modeling the dynamic events that occur in bio-tissue and characterizing viscoelasticity effects. In fact, it represents a tunable and more predictive modeling tool that is more adapted to the physical nature of the bio-materials \cite{jaishankar2013power, kobayashi2012modeling, kilbas2006theory}. Hence, fractional-order constitutive laws provide more stable and a more realistic tool than integer constitutive laws to simulate the viscoelastic materials, with fewer parameters to identify \cite{hollkamp2018model,magin2006fractional}. 
 
Here our goal is to investigate the use of fractional-order lumped element models to evaluate the aortic input impedance. It has been considered that the dynamics of the viscoelastic vessel wall can be more accurately displayed, using a fractional-order capacitor. This  non-ideal electrical element combines both the resistance and the capacitance that display the viscoelastic behavior of the systemic arteries. The fractional differentiation order  controls the contribution of both the resistance and the capacitance which allows an accurate and real physiological description \cite{bahloul2018three,bahloul2019two}. In our study we used tools from fractional-order calculus as well as two fractional-order lumped parametric models, the fractional-order two-element Windkessel $\mathrm {(FWK2)}$ and the fractional-order three-element Windkessel $\mathrm {(FWK3)}$ investigated to represent the main mechanical properties of the arterial system.
 The proposed models offer a a new, innovative tool to  better understand the viscoelastic effect and its resistive behavior on the motions of arteries. Another main goal of our study is to show that such a fractional-order model smoothly incorporates the complex effects and multi-scale properties of vascular tissues, using a reduced-order viscoelastic approach that provides a good estimation of the complex and frequency-dependent arterial compliance.
 To perform both quantitative and qualitative evaluations of the proposed fractional-order models, we investigated the in-silico ascending aortic blood pressure and flow database of 3,325 virtual subjects created by Willemet et al. \cite{willemet2015database}. We compared the proposed models with the inferred aortic input impedance.
Additionally, we performed detailed comparisons of the capabilities of the proposed models, the classical Windkessel: the two-element Windkessel $\mathrm {(WK2)}$ and the three-element Windkessel $\mathrm {(WK3)}$ models, for representing the in-silico aortic input impedance in various physiological states (normotensive, hypertension, severe-hypertension). Results showed that the proposed models capture very well the real dynamic and could monitor changes in the aortic input impedance modulus for different physiological states. Therefore, we can postulate that the fractional order lumped parametric modeling is a suitable candidate for solving the “hemodynamic inverse problem”. 

This paper is organized as follows: in Section II, the preliminary background is presented; the proposed model, the virtual database, and the parameters estimation method are discussed in Section III. Section IV shows the results of parameter calibration and further discussion. Finally, section V presents the conclusion and future perspectives.

\section{ Preliminary Background}
\subsection{Aortic Input Impedance}
Physiological investigations have established that the central blood pressure at the level of aorta depends on the properties of both the arterial system tree and the heart. Conventionally, investigators have used the concept of input impedance $ {(Z_{in})}$ in order to characterize the arterial network,  independently of the heart properties \cite{vlachopoulos2011mcdonald, noordergraaf2012circulatory, milnor1989vascular}. Hence, the aortic input impedance is considered to be a simple and complete descriptor of the arterial system, serving as the heart after-load. It provides a source of phenomenological information that only  depends on the geometrical and mechanical characteristic of the arterial tree and the blood it contains. $ {Z_{in}}$ is defined as a linear time invariant transfer function that links the arterial blood pressure $ {(P_a)}$ to the blood flow $ {(Q_a)}$ in the frequency domain. 
\begin{equation}
{Z_{in}}(\omega)=\frac{P_a(\omega)}{Q_a(\omega)}
\end{equation}
Where $ {\omega}$ is the frequency and $ {Z_{in}}$ is a complex function that comprises both real and imaginary parts. The hemodynamic analyses are usually based on the magnitude and phase of $ {(Z_{in})}$. Sometimes, investigators are interested in the quantification of the arterial network energy dissipation, hence they focus on the real value of $ {(Z_{in})}$, ${(Re [Z{in}])}$, which is considered, in this case, more appropriate for the evaluation of the arterial after-load \cite{quick2001constructive}. 

\subsection {Arterial Windkessel Model}
Windkessel models are  lumped parameter models, widely used  to characterize the vascular and pulmonary arterial systems. The historical Windkessel is the simplest mathematical model that can describe the overall function of human arterial systematic trees, using simple parameters such as total arterial compliance $ {(C)}$, aortic characteristic impedance $ {(Zc)}$ and total peripheral resistance $ {(Rp)}$. The main purpose of WK methods is to model the aortic input impedance as an after-load to the heart \cite{aboelkassem2019hybrid,westerhof2009arterial}.

 Conventionally, it is very convenient to adopt the electrical analogy, when using $\mathrm {WK}$, where the resistance, compliance and blood inertia are presented by a resistor, capacitor and inductor respectively. Additionally, the blood pressure corresponds to the electrical voltage and the flow rate to the current. Hence, the aortic input impedance is also equivalent to the electrical impedance expressed as the voltage to the current ratio, in the frequency domain \cite{aboelkassem2019hybrid}. 

The WK concept have been, firstly, introduced with the pioneer two-element Windkessel (WK2) by the German physiologist O. Frank in 1899, that simply described the whole arterial network in terms of capacitor connected in parallel to a resistor accounting respectively for the total arterial compliance and peripheral resistance. Fig.1 (a) shows the model analog circuit of $\mathrm {WK2}$.  The elementary $\mathrm {WK2}$ is simple, however it fails to estimate the aortic pressure, during the systolic phase of the cardiac cycle \cite{westerhof2009arterial, stergiopulos1995evaluation}.
\begin{figure}[!t]
	\centering
	\includegraphics[height=4cm,width=8.5cm]{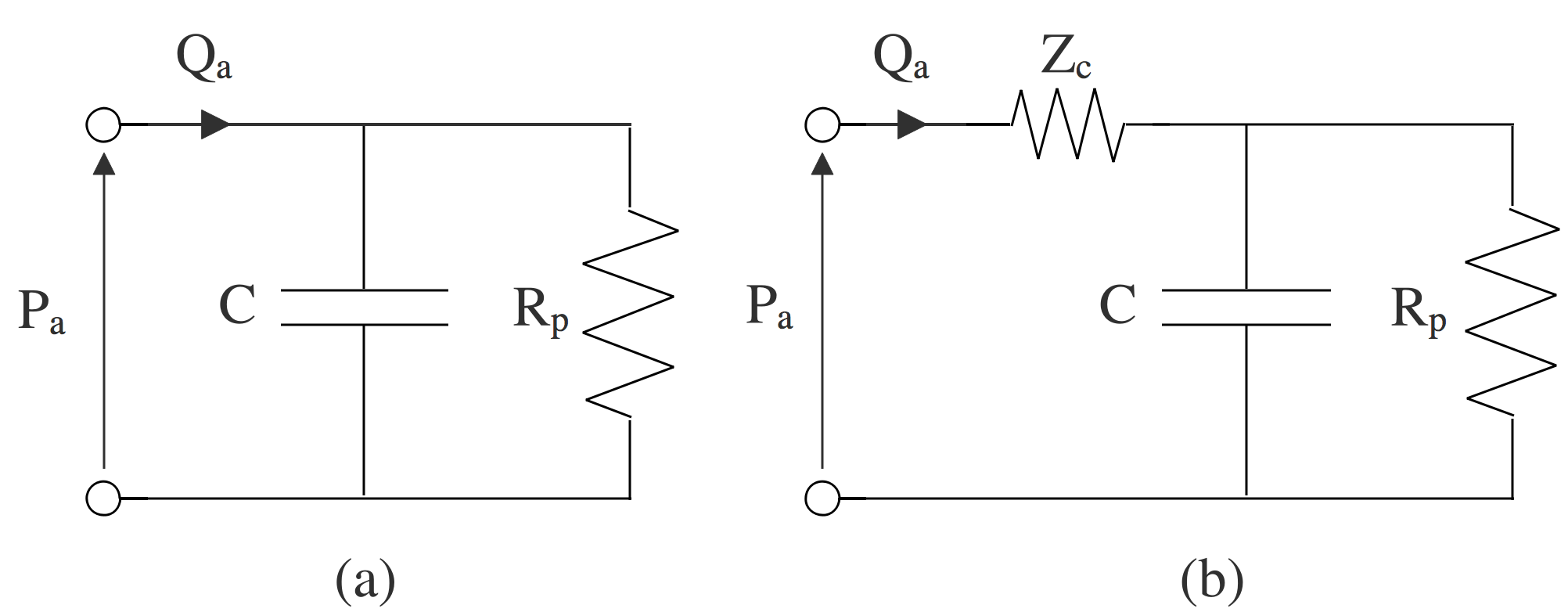}%
	\caption{ (a) Two-element Windkessel (WK2) analog representation. It simply describes the whole arterial system in terms of capacitor $(C)$ connected in parallel to a resistor $(R_P)$ accounting respectively for the total arterial compliance and peripheral resistance. (b) Three-element Windkessel analog representation that consists of a resistance $Z_C$ accounting for the aortic characteristic impedance, connected in series to WK2 model. }
	\label{fig1}
		\vspace{2mm}
\end{figure}
\setlength{\textfloatsep}{0cm}
To overcome the limitations of the WK2, a three-element Windkessel $\mathrm {(WK3)}$ model  was  introduced where  a characteristic impedance $ {Z_C}$  is combined to the $\mathrm {WK2}$ cell in series, (Fig. 1 (b)). This characteristic impedance provides a better description of the arterial input impedance at all frequency ranges \cite{capoccia2015development}. 

Several variants of the Windkessel model have been proposed to capture the arterial hemodynamics. These models vary by the  number of  electrical elements involved \cite{shi2011review}. Despite the progress made,   the compliance $ {C}$, that describes the elasticity  of the blood vessel,  is still not well represented. This is due t the fact that Windkessel models assume the arterial vessels to be purely elastic and thus where an ideal capacitor is usually used.  This contradicts the recent research work, which  explains that  the tissues of the arterial vessels are  viscoelastic similar to any bio-material \cite{burattini1998complex}.
\subsection{Arterial Viscoelasticity}
\subsubsection{Integer-order viscoelasticity arterial modeling}
The bio-mechanical property of the artery is defined  by its three main characteristics of the wall: smooth muscle fibers, elastin fibers, and collagen fibers. Like other soft collagenous tissue, it has long been recognized that arterial tissues exhibit a viscoelastic behavior, rather than a purely elastic one. In fact, the mechanical energy transported to the vessel wall is split into two parts: the stored and dissipated fractions. The first part is stored in a reversible form 
 to the elasticity property of the vessel, whereas the other part is dissipated due to its viscosity attribute \cite{reesink2018constitutive}. Generally, the arterial viscoelasticity is characterized using stress-relaxation experiments, and in order to capture its feature and fit such parameters, ordinary lumped element models based on integer-order differential equation are employed \cite{wang2016viscoelastic}.

Over the last century, a large number of integer-order models have been proposed to describe the stress-strain relationship. Conventionally, these models have used a mechanical analogy by connecting loss-less elastic elements (spring) and lossy viscous damper (dashpot) to represent respectively the elastic (often Hookean) and viscous (often Newtonian) properties of the bio-tissue. The Voigt mechanical cell (a spring connected in parallel to a dashpot), as shown in Fig. 2 (a), is the elementary model that can describe the viscoelastic behavior of the bio-mechanic collagenous tissue. With respect to the Windkessel modeling paradigm, some research attempts have been made to characterize the arterial viscoelasticity by developing a Viscoelastic Windkessel $\mathrm {(VWK)}$ configuration. $\mathrm {VWK}$ is similar to the standard $\mathrm {WK}$: however, in this configuration, the ideal constant capacitor $ {C}$ accounting for the total arterial compliance is substituted with a complex frequency-dependent capacitor $ {C_c(j\omega)}$. $ {C_c}$ corresponds to the electrical analogy of the standard Voigt cell. It contains a resistor $ {R_d}$ in serial with a capacitor $ {C_{vw}}$ representing respectively the viscous losses and static compliance (Fig. 2 (b) and (c)) \cite{capoccia2015development}. The complex frequency-dependent compliance can be expressed as follows: 
\begin{equation}
C_c(j\omega)=C_{vw}\frac{1}{1+jwR_dC_{vw}}.
\end{equation}
\begin{figure}[!t]
	\centering
	\vspace{0mm}
	\includegraphics[height=4cm,width=8.5cm]{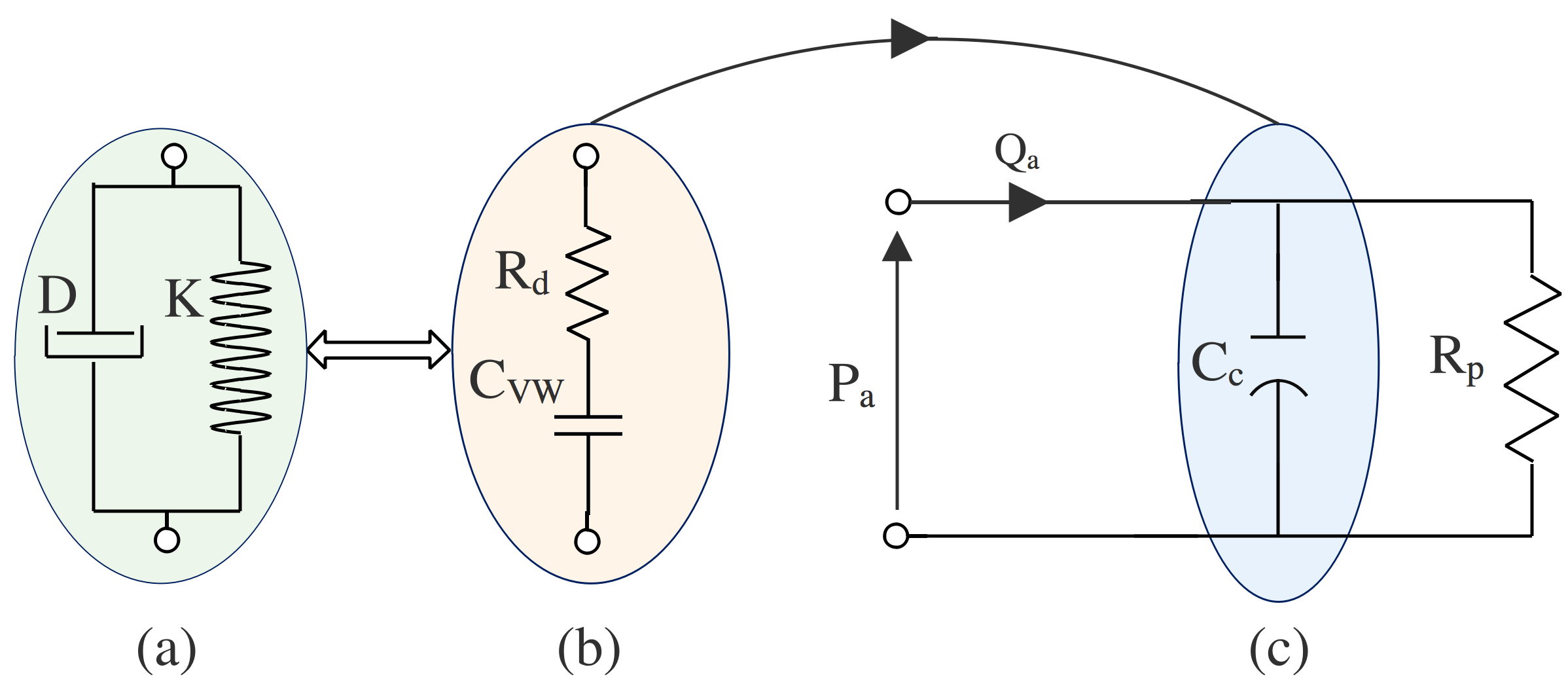}%
	\caption{ Mechanical representation of Voigt Model, (b) Equivalent electrical representation of the Voigt Model, (c) Electrical Voigt element based Viscoelastic Windkessel Model.}
	\label{fig2}
		\vspace{2mm}
\end{figure}
\setlength{\textfloatsep}{0cm}
 The use of Voigt cell to model the arterial vessel motion  is not accurate and present several limitations. The main reason for that is that the stress-relaxation property is not taken into account with  Voigt model. To address this inconsistency, researchers have suggested increasing the order of the viscoelastic representation by increasing the number of viscous and elastic connected lumped elements \cite{burattini1998complex}. These alternatives provide a more accurate but complex configuration. In fact, high-order viscoelastic models would endure to a more natural and realistic representation, however, their complexity and very large number of unknown parameters lead to a parameter identification issues. In general, simple models of reduced order are  desirable \cite{hollkamp2018model,naghibolhosseini2018fractional}. 
\subsubsection{Fractional-order viscoelasticity arterial modeling}
 Over past few years, numerous studies have shown that the arterial viscoelasticity can be well described using fractional-order  models.
 Previous research has shown the benefits of fractional-order systems over integer-order counterparts. For instance, modeling biological systems within a fractional-order framework allows investigators to reduce the order of complexity of the system and further improve the accuracy of estimated values of such parameters \cite{magin2006fractional}. With respect to the characterization of vascular system, it has been proven that due to the viscoelastic nature of the arterial walls, it is more adequate to describe the mechanical vascular properties using a fractional-order differential equation \cite{ref7,ref8, ref9, ref10}. 
 
 Accordingly, mechanical fractional-order viscoelasticity models $\mathrm {FVM}$ such as the Voigt-FVM and Standard-Linear-Solid-FVM have been used to investigate the effects of such hemodynamic index e.g., heart rate (HR)  on arterial viscoelasticity \cite{xiao2017arterial}. In general, $\mathrm {FVM}$ contains a pure spring and one or two fractional-order mechanical components (e.g., spring-pots). In this models, the fractional element displays the fractional-order derivative relationship between the mechanical stress $ {(\sigma(t))}$ and strain $ {(\epsilon (t))}$ on the vessel as described as follows:
 \begin{equation}
 \sigma(t)=\eta D^\alpha_t \epsilon(t)
 \end{equation}
 where $ {\alpha}$ is the fractional differentiation order parameter that controls the level of  viscoelasticity of the  artery, and $ {\eta}$ is a constant of proportionality.
  As $\mathrm {\alpha}$ approaches to $\mathrm {1}$, the artery's behavior is similar to a pure viscous dashpot (more resistive), and when $\mathrm {\alpha}$ border on $\mathrm {0}$, the vessel wall motion is more similar to that of a pure elastic spring. 
  
  With respect to the lumped parametric Windkessel paradigm we find that a key missing item is a fractional-order analog component that can display the real arterial viscoelastic behavior and combine its effects within Windkessel configurations.
 \subsection{Fractional-Order Derivatives}
 The concept of fractional-order derivative $\mathrm {(FD)}$ extends the conventional integer derivative to a non-integer order \cite{podlubny1998fractional}.   $\mathrm {FD}$ has attracted many researchers  from both theory and application  thanks to its  interesting properties such as its  non-locality and memory properties. $\mathrm {FD}$ is a powerful operator for modeling complex physical systems  in several fields of science and engineering, including biomedical  systems. It provides deep physiological insights and introduces  new parameters that can capture the overall behavior of a system with fewer equations comparing to its integer order models counterparts. In addition,  recent studies have demonstrated the capability of  FD to describe accurately  the viscoelasticity properties of biological tissues \cite{freeborn2013survey}.  
 
 The integral and the differential operators  have been generalized into an unified (differ-integration) operator $ {D_t^\alpha}$,  introduced as:
 \begin{equation}
 D_{\alpha}^{t}=
 \left\{\begin{matrix}\frac{\mathrm{d^{\alpha}} }{\mathrm{d} t^{\alpha}} 
 & if & \alpha>0 \\ 
 1,& if & \alpha=0\\ 
 \int_{t}^{0}\left ( df \right )^{-\alpha}& if & \alpha<0
 \end{matrix}\right.
 \end{equation}
 where ${\alpha}$ is an arbitrary real order of the operator (integral or derivative) known as the fractional order, and $ {df}$ is the derivative function. Numerous fractional calculus definitions have been suggested. Generally, these definitions can be classified into two main classes. In the first class, the operator $ {D_t^\alpha}$ is converted into the standard differential-integral operator when $ {\alpha}$ is integer. For instance, the $\mathrm {Reimann}$-$\mathrm {Liouville}$ definition of a fractional-order derivative $ {\alpha}$ of a function $ {g(t)}$ is given by  \cite{jumarie2006modified}:
 \begin{equation}
 D_t^{\alpha}g\left ( t \right )=\frac{1}{\Gamma \left ( 1-\alpha  \right )}\frac{\mathrm{d} }{\mathrm{d} t}\int_{0}^{t}\frac{g\left ( \tau  \right )}{\left ( 1-\tau  \right )^{\alpha }}d\tau, 
 \end{equation}
 where $ {\Gamma}$ is the Euler gamma function. The second class is that the Laplace transform of $ { D_t^\alpha}$  is $ {s^\alpha}$, assuming null initial fractional conditions. The fractional operator is given by:
 \begin{equation}
 D_t^{\alpha}g\left ( t \right )\overset{L}{\rightarrow}s^{\alpha}G\left ( s \right ),
 \end{equation}
 This class is very interesting in developing parametric models for complex systems and control design in frequency domain. $ {s}$ is the complex Laplace. The Fourier transform can be found by substituting $ {s}$ by $ {j\omega}$ and thus the equivalent frequency-domain expression of  $ {s^\alpha}$: 
  \begin{equation}
  \left ( j \omega \right )^\alpha=\omega^\alpha\left ( cos\frac{\alpha\pi}{2} -jsin\frac{\alpha\pi}{2}\right ),
  \end{equation}
  \begin{equation}
  \frac{1}{\left ( j\omega \right )^\alpha}=\frac{1}{\omega^\alpha}\left ( cos\frac{\alpha\pi}{2} +jsin\frac{\alpha\pi}{2}\right ).
  \end{equation}
\subsection{Fractional-order capacitor }
$\mathrm {FoC}$ is usually defined as a  constant phase element $\mathrm {(CPE)}$. It generates an   impedance with  phase angle  between $\mathrm {0^{\circ}}$ and $\mathrm {-90^{\circ}}$ for all frequency ranges  \cite{ref21}\cite{ref22}. $\mathrm {FoC}$ offers exceptional advantages for impedance modeling by enabling a large impedance matching range and allowing the flexibility of frequency response of such system. Mathematically, the relationship between the current $ {(i_c)}$ across $\mathrm {FoC}$ and its voltage $ {(v_c)}$ in the time domain is given by:  
\begin{equation}
i_{c}(t)= \frac{1}{A_\alpha} \  D_t^{\alpha}\ v_c(t).      
\end{equation}
Applying Laplace transformation as defined in (6) to (9) and assuming null initial conditions we obtain:
\begin{equation}
{Z_{F}}\left ( s \right )=\frac{V_c(s)}{I_c(s)}= A_\alpha\ s^{-\alpha },
\end{equation}
where $ {{A_\alpha}} $ is known as the coefficient of the pseudo-capacitance expressed in units of $\mathrm {Farad.sec^{\alpha-1}}$, and $\mathrm {\alpha}$ is the fractional differentiation order. It is worth noting that the fractional impedance is variable, with an exponent $\mathrm {\alpha}$ $\mathrm {\left ( 0< \alpha <1  \right )}$. By substituting $\mathrm {s}$ with $\mathrm {j\omega}$, where $\mathrm {j}$ is a complex number and $\mathrm {\omega}$ is the radial frequency, the complex value of $\mathrm {FoC}$, at a specific frequency $ {\omega}$ can be obtained using the following expression:
\begin{equation} 
{Z_{F}}\left ( jw \right )=  \underbrace{A_\alpha w^{-\alpha } cos(\phi)}_\text{\large$Z_D$}+j\underbrace{A_\alpha w^{-\alpha }sin(-\phi)}_\text{\large $Z_S$},
\end{equation}
where $ {\phi\!=\!\alpha \frac{\pi}{2}}$. For $ {\alpha\!=\!0}$, $ {Z_{F}}$ refers to an ideal resistor with a phase angle of $\mathrm {0^{\circ}}$. For $\mathrm {\alpha\!=\!1}$, $ {Z_{F}}$ remains an ideal capacitor with a phase angle equal to $\mathrm {-90^{\circ}}$. 
 The modulus at a specific radial frequency $\mathrm {\omega}$ can be expressed as: 
\begin{equation}
\left | Z_{F}\right |=\sqrt{\left ( A_\alpha w^{-\alpha }cos\left ( \phi \right )\right )^{2}\!\!+\left ( A_\alpha w^{-\alpha }sin\left ( -\phi \right )\right )^{2}}.
\end{equation}
\begin{figure}[!t]
	\centering
	\includegraphics[height=3cm,width=11cm]{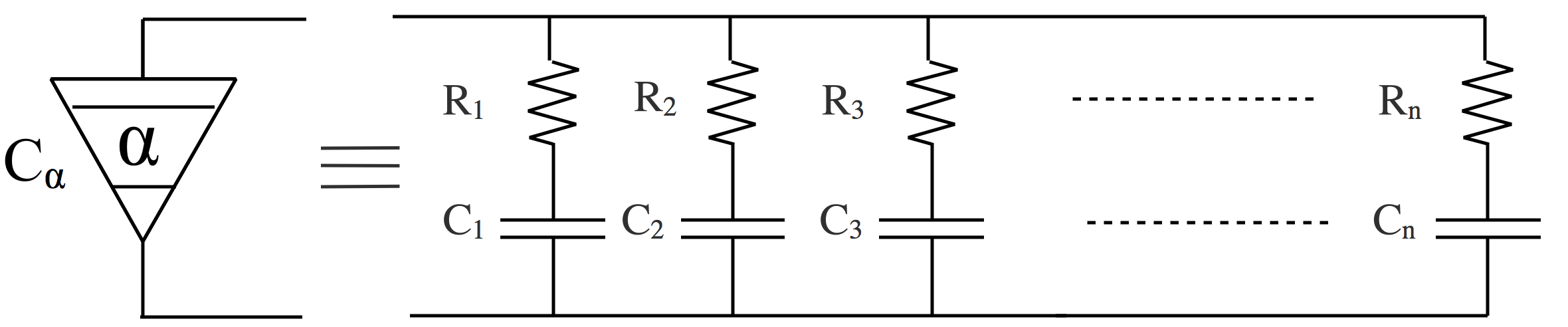}%
	\caption{ RC network for fractional order capacitor emulation.}
	\label{fig3}
		\vspace{2mm}
\end{figure}
\setlength{\textfloatsep}{0mm}
In the past, researchers have tried to approximate $\mathrm {FoC}$ by combining conventional  electrical components. Accordingly, $\mathrm {FoC}$ behavior can be generated using the association of resistors and capacitors  as shown in Fig. 3 \cite{ref29}. The resulting network is analogue to the equivalent electrical circuit  of n-serial Voigt mechanical cells and is also equivalent to the generalized Kelvin-Voigt viscoelastic model. Furthermore, From (11), $ {Z_{F}}$ is decomposed into a dissipative term $ {Z_{D}}$ (real part) and a storage term $ {Z_{S}}$ (imaginary part).  Using mechanical analogy,  these terms might be used to represent the viscous and elastic behaviors of the arteries. 
In this paper, the proposed modeling approach uses a $\mathrm {FoC}$  which account for the viscoelastic property of the vessel motion. The fractional order capacitor  includes  both resistive and capacitive characteristics that generate  the  viscoelastic behavior of the arteries. The contribution of each property is controlled  by the fractional differentiation parameter  $\mathrm {\alpha}$, which provides a reduced and flexible description of the underling physiological phenomena.
\section{Methods}
In this section, we introduce two fractional-order models, $\mathrm {FWK2}$ and $\mathrm {FWK3}$, that characterize the main mechanical arterial properties of the arterial system. The proposed models belong to the standard lumped parametric Windkessel; however, the order $\mathrm {\alpha}$ of these models is fractional $\mathrm {(\alpha\! \in \!\mathbb{R} \ (n<\alpha<n+1))}$ rather than integer $\mathrm {(n\in \mathbb{N})}$. 
\begin{figure*}[!t]
	\centering
	\includegraphics[height=11cm,width=17.7cm]{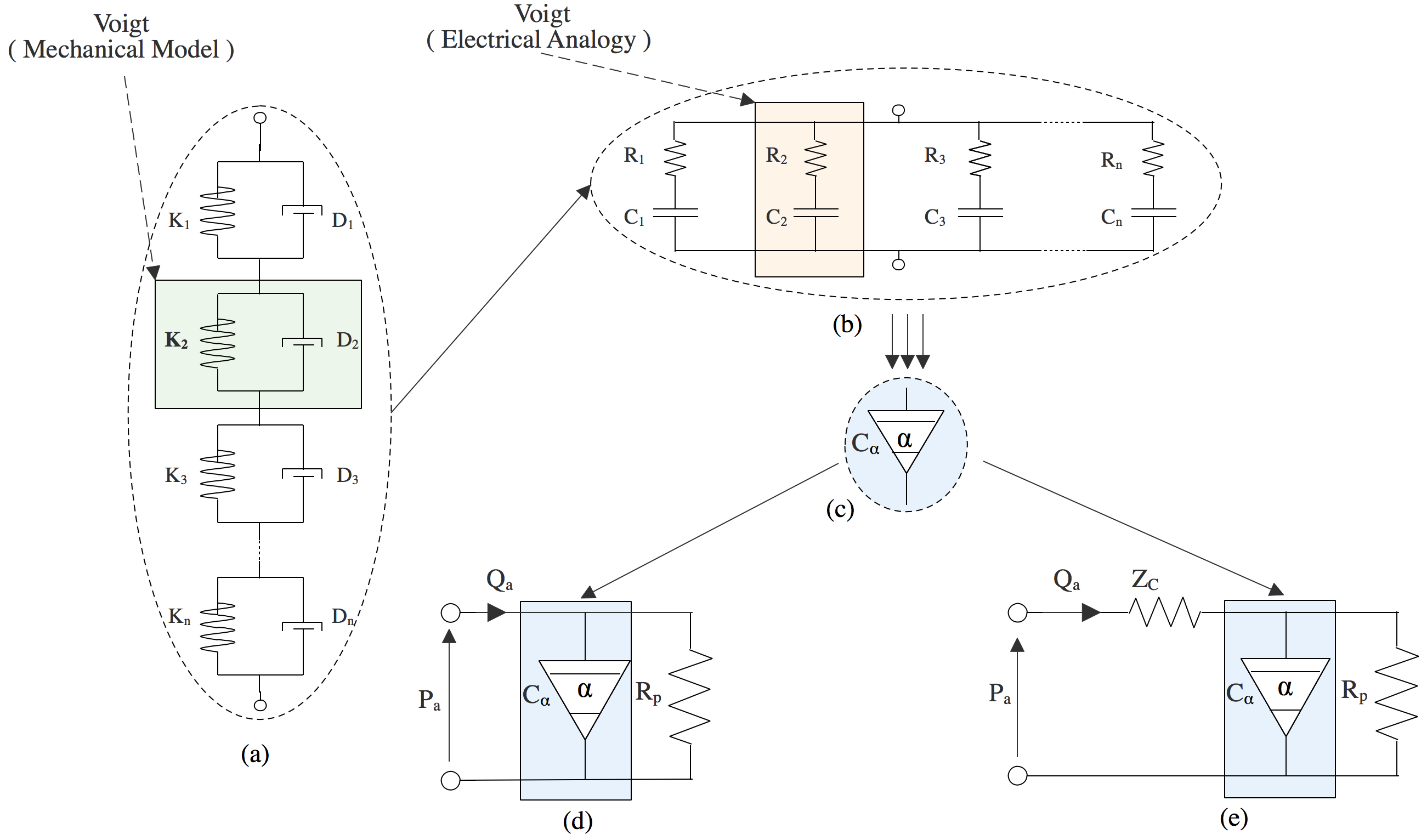}%
	\caption{ (a) Generalized Kelvin-Voigt viscoelastic model consisting of $n$ serial Voigt Cells, (b) Electrical analogy of the generalized Kelvin-Voigt that is equivalent to a ladder network comprising $n$ ($RC$) branches connected in parallel, (c) Fractional-order capacitor $C_\alpha$ representing the fractional-order compliance where $\alpha$ is the fractional differentiation order, (d) Fractional-order two-element Windkessel model analog circuit, and (d) Fractional-order three-element Windkessel model analog circuit.}
	\label{fig3}
	\vspace{2mm}
\end{figure*}
 Fig. 4 illustrates the concept of viscoelastic model reduction using $\mathrm {FoC}$. In fact, by examining the electrical analogy between the generalized Kelvin-Voigt viscoelastic model (Fig. 4 (a)) and the equivalent circuit of the $\mathrm {FoC}$ (Fig. 4 (b)), we can clearly notice that they are analogous. Bearing this in mind, we replace the constant capacitor $\mathrm {(C)}$ representing the total arterial compliance by $\mathrm {FoC}$ $\mathrm {(C_\alpha)}$ in both $\mathrm {WK2}$ and $\mathrm {WK3}$. When the fractional differentiation order $\mathrm {(\alpha)}$ is equal to one, the model then describes the special case of the standard integer-order morels i.e ($\mathrm{WK2}$ and $\mathrm {WK3}$). Hence, we show only the derivations of $\mathrm {FWK2}$ and $\mathrm {FWK3}$. 
\subsection { Fractional-order two-element Windkessel (FWK2)}
Similarly to the conventional $WK2$, the proposed model consists of a  $\mathrm {FoC}$ $ {(C_\alpha)}$, in parallel with a resistor $ {R_p)}$ representing  the total complex and frequency-dependent total arterial compliance and the arterial peripheral resistance, respectively.  The application of $\mathrm {Kirchhoff’s\ Current\ Law}$ on  the fractional circuit shown in Fig. 4 (d), will give:
\begin{equation}
Q_{a}(t)=C_{\alpha} D^{\alpha}_{t}P_{a}\left ( t \right )+\frac{P_{a}\left ( t \right )}{R_{p}}.
\end{equation}
Thus, the fractional aortic input impedance in the frequency domain can be obtained by applying the Laplace transformation to (13) and assuming null initial conditions:
\begin{equation}
{Z}_{2}^{\alpha}(s)=\frac{R_p}{1+\left(\tau_\alpha s\right)^\alpha},
\end{equation}
where 
\begin{equation}
\tau_\alpha=\sqrt[\alpha]{R_pC_{\alpha}}.
\end{equation}
The fractional aortic input impedance modulus and phase angle at a specific radial frequency $ {\omega}$ can be calculated using these formulas:
\begin{equation}
\left | {Z}_{2}^{\alpha}\right |\!=\!\!\frac{R_p}{\sqrt{[1\!+\!(\omega \tau_\alpha)^{\alpha } cos(\alpha\frac{\pi}{2})]^{2}\!+\![(\omega \tau_\alpha)^{\alpha } sin(\alpha\frac{\pi}{2})]^{2}}},
\end{equation}	

\begin{equation}
\angle {Z}_{2}^{\alpha}=-tan^{-1}\left ( \frac{(\omega \tau_\alpha)^{\alpha }sin(\alpha\frac{\pi}{2})}{1+(\omega \tau_\alpha)^{\alpha } cos(\alpha\frac{\pi}{2})} \right ).
\end{equation}

\subsection{Fractional-order three-element Windkessel (FWK3)}
 Fig 4. (e) illustrates the proposed model's scheme. The $Kirchhoff’s\ Voltage\ law$ lead to the following equation :
\begin{equation}
V_{C_{\alpha}}(t)=P_{a}(t)-Z_{C}Q_{a}(t)
\end{equation}
The application of the  ${Kirchhoff’s\ Current\ law}$ gives:
\begin{equation}
Q_{a}(t)=C_{\alpha}D_{ t}^{\alpha}V_{C_{\alpha}}(t)+\frac{V_{C_{\alpha}}(t)}{R_{p}}.
\end{equation}
Substituting (18) in (19) gives the following equations:

	\begin{equation}
	Q_{a}(t)\!\!=\!\!C_{\alpha}D_{t}^{\alpha}\left (\! P_{a}(t)\!\!-\!\!Z_{C}Q_{a}(t) \right )+\frac{\left ( P_{a}(t)\!\!-\!\!Z_{C}Q_{a}(t) \right )}{R_{p}},
	\end{equation}
\begin{equation}
Z_{C}C_{\alpha}D_{t }^{\alpha}Q_{a}(t)\!+\!Q_{a}(t)\left (\! 1\!+\!\frac{Z_{C}}{R_{p}}\! \right )\!\!=\!\!C_{\alpha}D_{t }^{\alpha}P_{a}(t)\!+\frac{P_{a}(t)}{R_{p}}.
\end{equation}

 Using Laplace transformation with the assumption of  null initial conditions will lead to the following  fractional-order aortic input impedance $ {{Z}_{3}^{\alpha}} $:
\begin{equation}
{Z}_{3}^{\alpha}(s)=Z_C+\frac{R_p}{1+\left(\tau_{\alpha} s\right)^\alpha},
\end{equation}
Or
\begin{equation}
{Z}_{3}^{\alpha}\left ( s \right )=(R_{p}+Z_{C})\frac{1+(\tau _{\alpha_{N}}s)^{\alpha }}{1+(\tau _{\alpha_{D}}s )^{\alpha }},
\end{equation}
where $ {\tau _{N}}$ and $ {\tau _{D}}$ expressed as follows:
\begin{equation}
\left\{\begin{matrix}
\tau _{N}=\sqrt[\alpha]{\frac{R_{p}Z_{C}}{R_{p}+Z_{C}}C_{\alpha}},\\ 
\\
\tau _{D}=\sqrt[\alpha]{R_{p}C_{\alpha}}.
\end{matrix}\right.
\end{equation}
Thus, the input impedance modulus and the phase angle, at a specific frequency $ {\omega}$, are given by (25) and (26):

\begin{equation}\tag{25}\normalsize
\left | {Z}_{3}^{\alpha} \right |=(R_{p} +Z_{C})\frac{\sqrt{(1+(w\tau _{N})^{\alpha }cos(\alpha\frac{\pi}{2}))^{2}+((w\tau _{N})^{\alpha }sin(\alpha\frac{\pi}{2}))^{2}}}{\sqrt{(1+(w\tau _{D})^{\alpha }cos(\alpha\frac{\pi}{2}))^{2}+((w\tau _{D})^{\alpha }sin(\alpha\frac{\pi}{2}))^{2}}}   .
\end{equation}

\begin{equation}\tag{26}\normalsize
\angle {Z}_{3}^{\alpha}=\tan^{-1}\left ( \frac{(w\tau _{N})^{\alpha }sin(\alpha\frac{\pi}{2})}{1+(\omega\tau _{N})^{\alpha }cos(\alpha\frac{\pi}{2}))} \right )-\tan^{-1}\left ( \frac{(w\tau _{D})^{\alpha }sin(\alpha\frac{\pi}{2})}{1+(w\tau _{D})^{\alpha }cos(\alpha\frac{\pi}{2}))} \right ).
\end{equation}  

\subsection{In-silico Virtual Population}
\begin{figure}[t]
	\centering
	\vspace{0mm}
	\includegraphics[height=9cm,width=11cm]{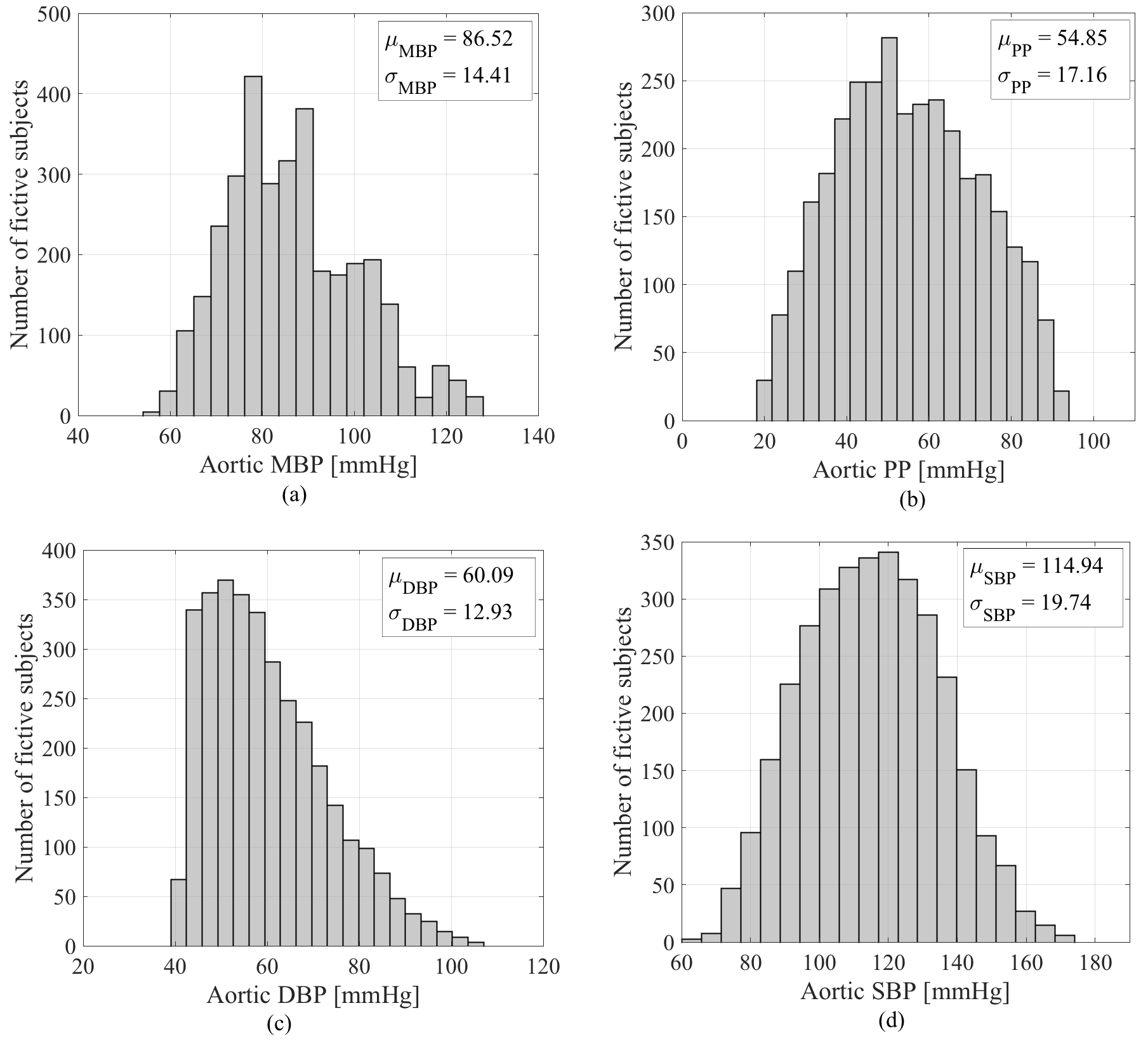}%
	\caption{ Distribution, mean value and standard deviation in (mmHg) of: (a) mean blood pressure (MBP), (b) pulse pressure (PP), (c) diastolic pressure (DP) and systolic pressure (SP) at the level of of ascending aorta for the in-silico data-base.}
	\label{fig3}
	\vspace{2mm}
\end{figure}
To validate the proposed models, we use the virtual blood pressure and the flow waveforms at the level of the ascending aorta. To do so, we used the data from a database of 3, 325 virtual healthy adult subjects previously created by Marie Willemet et. al  \cite{willemet2015database}, in-silico, from a validated one-dimensional numerical model of the arterial hemodynamics, in which cardiac and arterial parameters vary within physiological ranges.The model is able to generate the major hemodynamic properties sensed in-vivo.
\footnote{\textcolor{black}{\url{http://haemod.uk/virtual-database}}.} 
Fig.5 shows a summary statistic of the  aortic blood pressure at the level of the ascending aorta, for all virtual subjects. It is clear that this database presents physiological values with well-balanced distributions. The Cardiac outputs vary between $\mathrm {3.5}$ and $\mathrm {7.2 l/min}$, depending on the values of the heart rate $\mathrm {(53, 63, and 72 beats/min)}$ and stroke volume $\mathrm {(66, 83, and 100 ml)}$ prescribed.
\subsection{Model Calibration}
For both the classical $\mathrm {WK}$ and the proposed fractional-order models, the total peripheral resistance parameter, $ {R_p}$, was evaluated from the ratio of mean pressure to the mean blood flow ${(R_p=\frac{\overline{P_a}}{\overline{Q_a}})}$. The other parameters used in the proposed models, $ {\Theta_{Z_{2}^{\alpha}}=\{\tau_\alpha, \alpha\}}$ of $\mathrm {FWK2 }$, $ {\Theta_{Z_{3}^{\alpha}}=\{Z_C, \tau_{\alpha}, \alpha\}}$ of $\mathrm {FWK3 }$ and the classical Windkessel, $ {\Theta_{Z_{2}}=\{C\}}$ of $\mathrm {WK2 }$ and $ {\Theta_{Z_{3}}=\{Z_C, C\}}$ of $\mathrm {WK3}$, were estimated by fitting the models to in-silico data. We first calculated the in-silico aortic input impedance. To do so, the in-silico blood flow and pressure for a single cardiac cycle expressed in the time domain were converted to the frequency domain, using a discrete Fourier transform, and the input impedance was formulated as the ratio of the harmonic of the blood pressure to the flow. The optimal $\Theta_{Z_{2}^{\alpha}}$ (17), $\Theta_{Z_{2}}$ ((17) for $ {\alpha=1}$), $\Theta_{Z_{3}^{\alpha}}$ (22) and $\Theta_{Z_{3}}$ ( (22) for $ {\alpha=1}$),  were determined by solving the following optimization problem:
\begin{equation}\normalsize\tag{27}
\Theta^*=  \arg\min_{\Theta} f(\Theta),
\end{equation}

\begin{equation}\tag{28}
 {f(\Theta)\!\!=\!\!\sqrt{\!\frac{\sum_{i=1}^{N}\!\left \{\left[Re(Z_{[i]})\!\!-\!\!Re(\hat{Z}_{[i]}(\Theta))\right]^2\!\! +\!\!\left [Im(Z_{[i]}])\!\!-\!\!Im(\hat{Z}_{[i]}(\Theta))\right]^2 \right \}    } {N }}},
\end{equation}

where $ {\Theta^*}$ is the optimal $ {\Theta}$ minimizing the cost function (28), which corresponds to the root mean square error $\mathrm {(RMSE) }$. $ {Re}$ and $ {Im}$ denote the real and imaginary parts of the in-silico aortic input impedance $ {{Z}}$ and  the modeled impedance $ {\hat{Z}}$  evaluated at a specific harmonic $ {i}$. $ {N}$ is the total number of harmonics taken into account. 

To evaluate the performance of our model, the deviation of the model modulus from the in-silico aortic input impedance modulus was calculated, using the following expression:
\vspace{0mm}
\begin{equation}\tag{29}
D_i\ [\%]=\left [ \frac{\left | \hat {Z_{[i]}}(\Theta) \right |-\left| {Z_{[i]}} \right |}{\left| Z_{[i]} \right |}\right]_{i=1..N}100 \%.
\end{equation} 
For ease of visualization of the various comparisons between the different models, for each virtual subject, we evaluated the mean of D [\%] over the $N$ harmonics, based on the following equation:
\begin{equation}\tag{30}
Deviation\ [\%]=\frac{\sum_{i=1}^{N}D_{[i]} [\%]} {N}.
\end{equation}
Additionally, the normalized mean square error $\mathrm {(NMSE)}$ (31) was used to evaluate performance of our model when applied to the phase angle, using the following expression:
\begin{equation}\normalsize\tag{31}
NMSE=1-\frac {\left \| \angle {Z}-\angle \hat {Z}(\Theta)  \right \|^2}{\left \|\angle {Z}-mean (\angle  Z)  \right \|^2},
\end{equation}
where, $\mathrm {\|.\|}$  indicates the $\mathrm {2\!-\!norm}$ of a vector. $\mathrm {NRMSE}$ costs vary between $\mathrm {-Inf}$ (bad fit) to $\mathrm {1}$ (perfect fit).
\section{Results \& Discussion}
\begin{figure}[!t]
	\centering
	\vspace{0mm}
	\includegraphics[height=10cm, width=11cm]{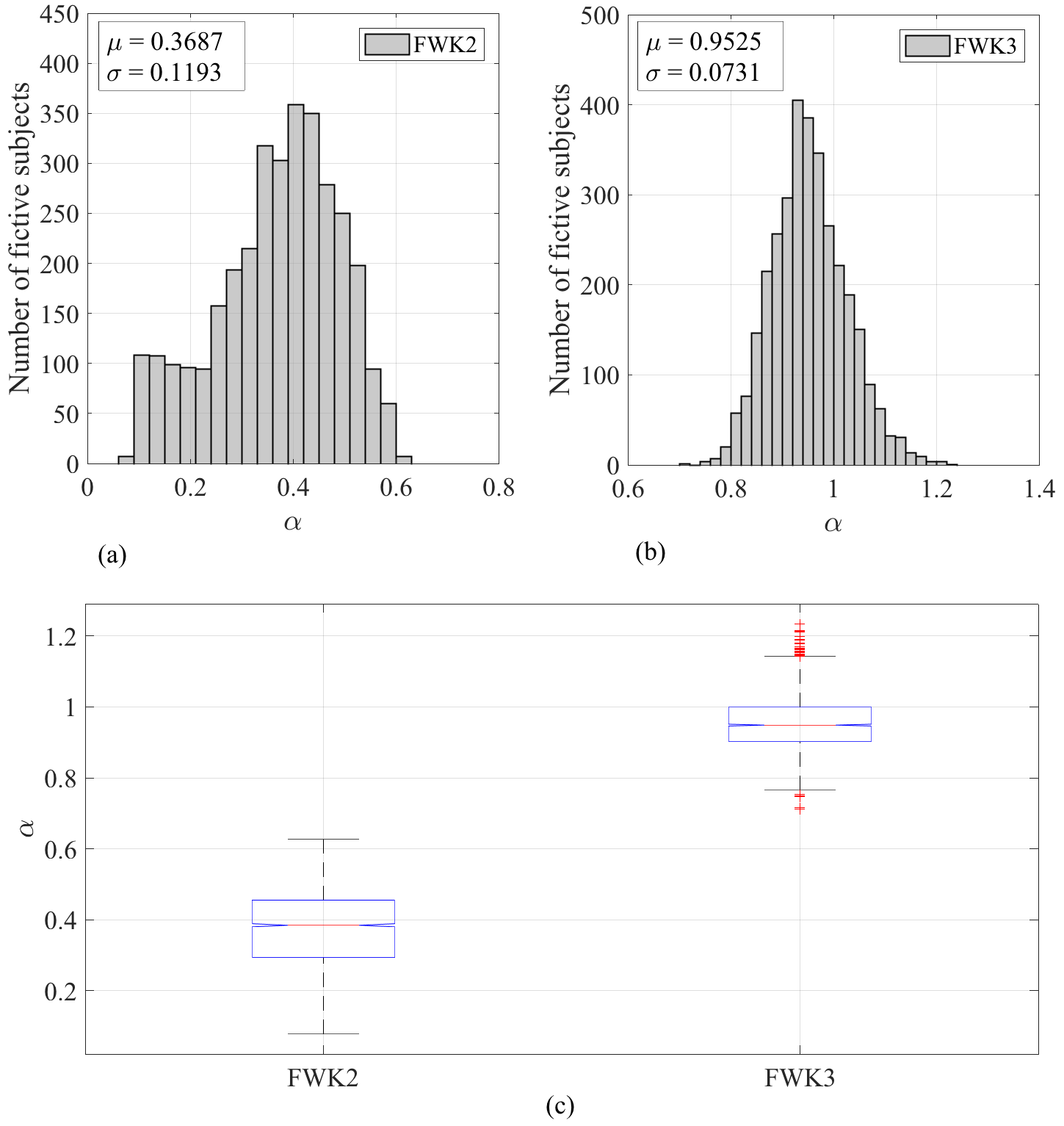}%
	\caption{Distribution of the estimates fractional differentiation order parameters $\alpha$ for FWK2 in (a) and  FWK3 in (b). (c) presents box plots, showing a visualization of summary statistics of a comparison between estimates $\alpha$ for the proposed models.}
	\label{fig3}
	\vspace{2mm}
\end{figure}
In this section, we present the results obtained for the in-silico data, using $\mathrm {FWK2}$ and $\mathrm {FWK3}$. To validate and check the ability of the proposed models to reconstruct the patterns of the in-silico aortic input impedance modulus and phase angle, we compare the results with those obtained with the classical $\mathrm {WK2}$ and $\mathrm {WK3}$. Parameters for the models $\mathrm {(WK2: \{\Theta_{Z_{2}}\},\ FWK2: \{\Theta_{Z_{2}^\alpha}\},\ WK3: \{\Theta_{Z_{3}}\}}$ and $\mathrm{ FWK3: \{\Theta_{Z_{3}^\alpha}\})}$ are determined numerically using the appropriate optimization technique implemented in $\mathrm {MATLAB\_R2018b}$ based on the cost functions defined in the previous section.
\begin{figure}[!t]
	\centering
	\vspace{0mm}
	\includegraphics[height=10cm, width=13cm]{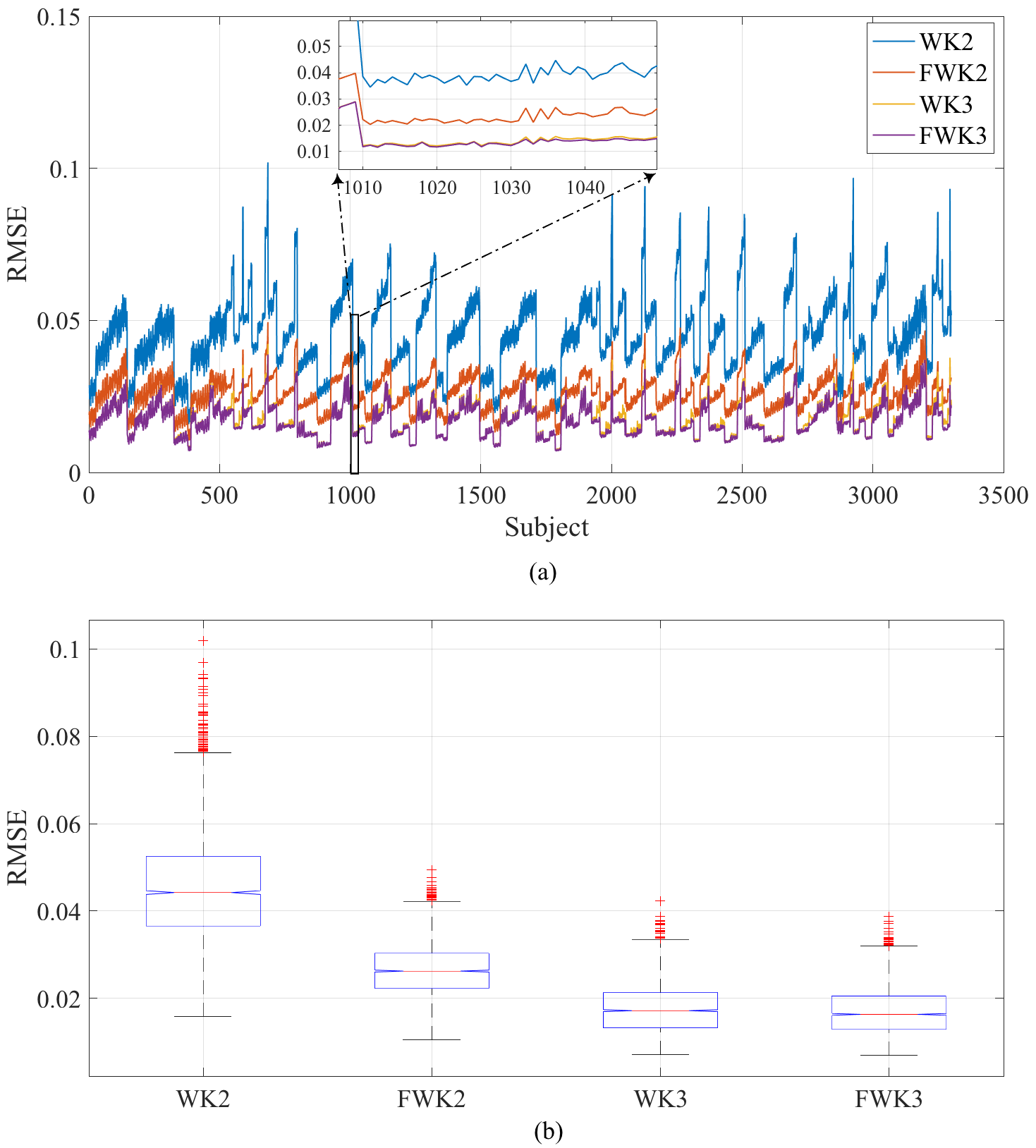}%
	\caption{(a) Comparison of goodness of fit quantified as $\mathrm {RMSE}$, evaluated for both proposed models (FWK2 and FWK3) and standard Windkessel (WK2 and WK3) for all the virtual subjects, and (b) box plots providing a visualization of summary statistics of this comparison.}
	\label{fig3}
	\vspace{2mm}
\end{figure}
\begin{figure}[!b]
	\centering
	\vspace{0mm}
	\includegraphics[height=10cm, width=13cm]{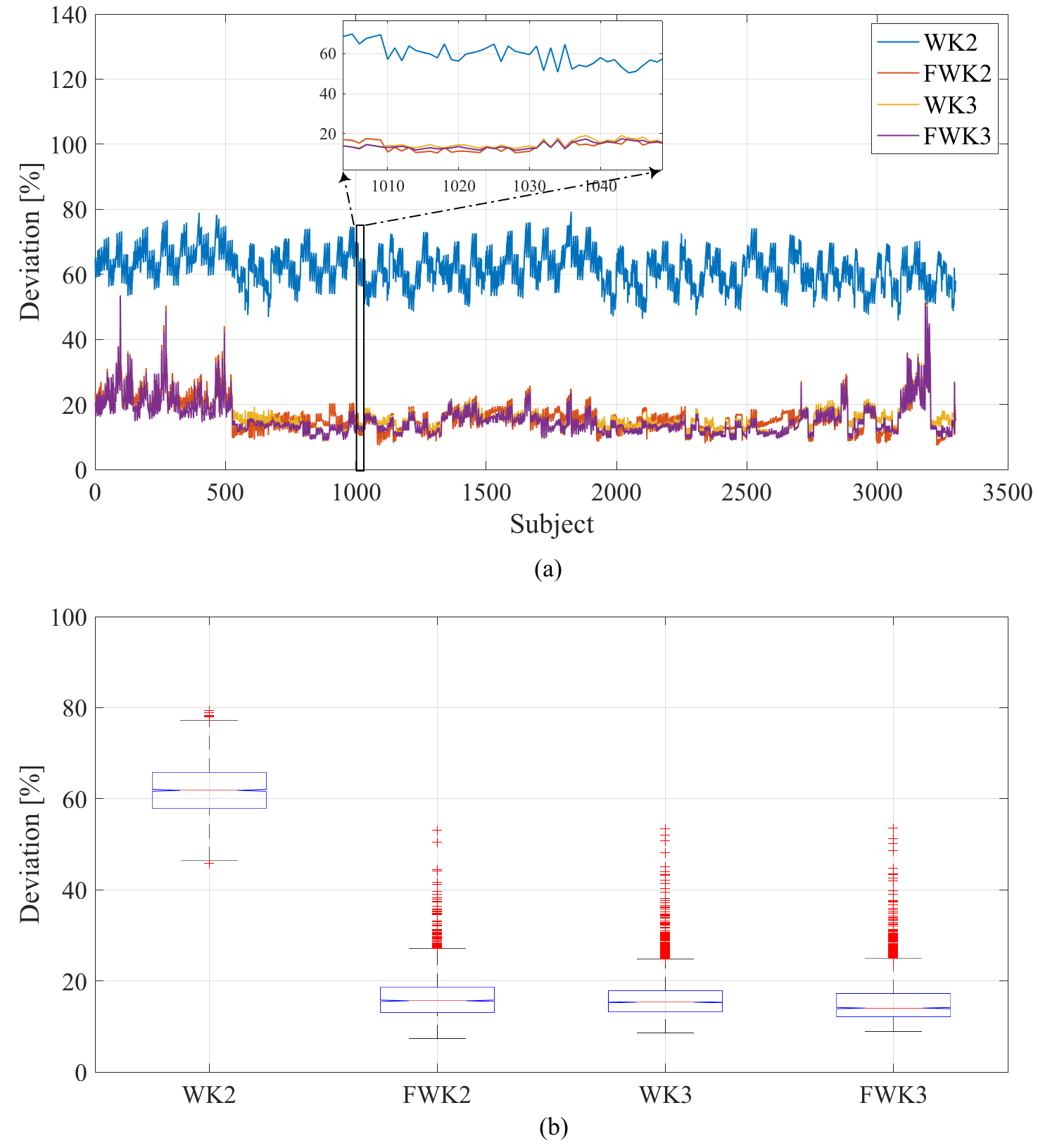}%
	\caption{(a) Comparison of goodness of fit of the aortic input impedance modulus quantified as $\mathrm {Deviation [\%]}$, evaluated for both proposed models (FWK2 and FWK3) and standard Windkessel (WK2 and WK3) for all virtual subject, and (b) box plots providing a visualization of summary statistics of this comparison.}
	\label{fig3}
	\vspace{2mm}
\end{figure}
 Table 1. presents the mean of $ {RMSE}$, $ {Deviation\ [\%]}$ and $ {NMSE}$ as quantifier of the overall performance of our model, including the modulus, the phase angles, and the arterial parameters.\
Fig. 6 (a) and (b) show the distribution of the estimated fractional differentiation order $ {\alpha}$, after fitting the proposed fractional-order impedance models $\mathrm {FWK2}$ and $\mathrm {FWK3}$ respectively, to the in-silico aortic input impedances evaluated at the level of the ascending aorta, for $\mathrm {3325}$ virtual subjects. In most of the subjects, $ {\alpha}$ values are different from the integer order (one) that corresponds to the order of classical Windkessel configurations. With respect to $\mathrm {FWK2}$, $\alpha$ is never equal to one, and for all the virtual subjects, its mean estimate value is approximately $0.3687 \pm 0.0081$. With respect to FWK3, the mean estimated value  of $\alpha$ is approximately $0.9525 \pm 0.005$. Additionally, for some subjects, the estimated value of $\alpha$ is above one. These results clearly indicate that the arterial systemic system exhibits a viscoelastic behavior, not a purely elastic one. Indeed, the fact that $ {\alpha\neq1}$ shows that the $\mathrm {FoC}$ element incorporates both resistive and capacitive quantities, as demonstrated mathematically in (11); it further supports the concept of a fractional-order behavior by the arterial system. In the proposed models, the fractional element, $ {C_\alpha}$, combines both resistive and capacitive properties and displays the viscoelastic behavior of the arterial vessel. The contributions from both properties are controlled by the fractional differentiation order $ {(\alpha)}$ enabling a more flexible physiological description. Due to this fact, it is easy to understand the difference between $ {\alpha}$ of $\mathrm {FWK2}$ and $ {\alpha}$ of $\mathrm {FWK3}$ (Fig. 6 (c)). As a result, $\alpha$ decreases from one to zero, the resistive part increases in $\mathrm {FoC}$. On another hand, for the $\mathrm {FWK3}$ model, we have added a small resistance characterizing the characteristic aortic impedance $ {(Z_C)}$ comparing to $\mathrm {FWK2}$. Hence, $\alpha$ increases towards one (i.e., the resistive part of $\mathrm {FoC}$ decreases) as a counteraction of the contribution of $Z_C$. Indeed, in the $\mathrm {FWK2}$ model the resistive part of the $\mathrm {FoC}$ represents the total resistance of the systemic system; however, in the case of $\mathrm {FWK3}$, this quantity is shared between the two lumped elements $Z_C$ and FoC ($C_{\alpha}$). This supports the idea that the systemic system has a viscoelastic fractional-order behavior. 
\begin{figure}[!t] 
	\centering
	\vspace{0mm}
	\includegraphics[height=10cm, width=13cm]{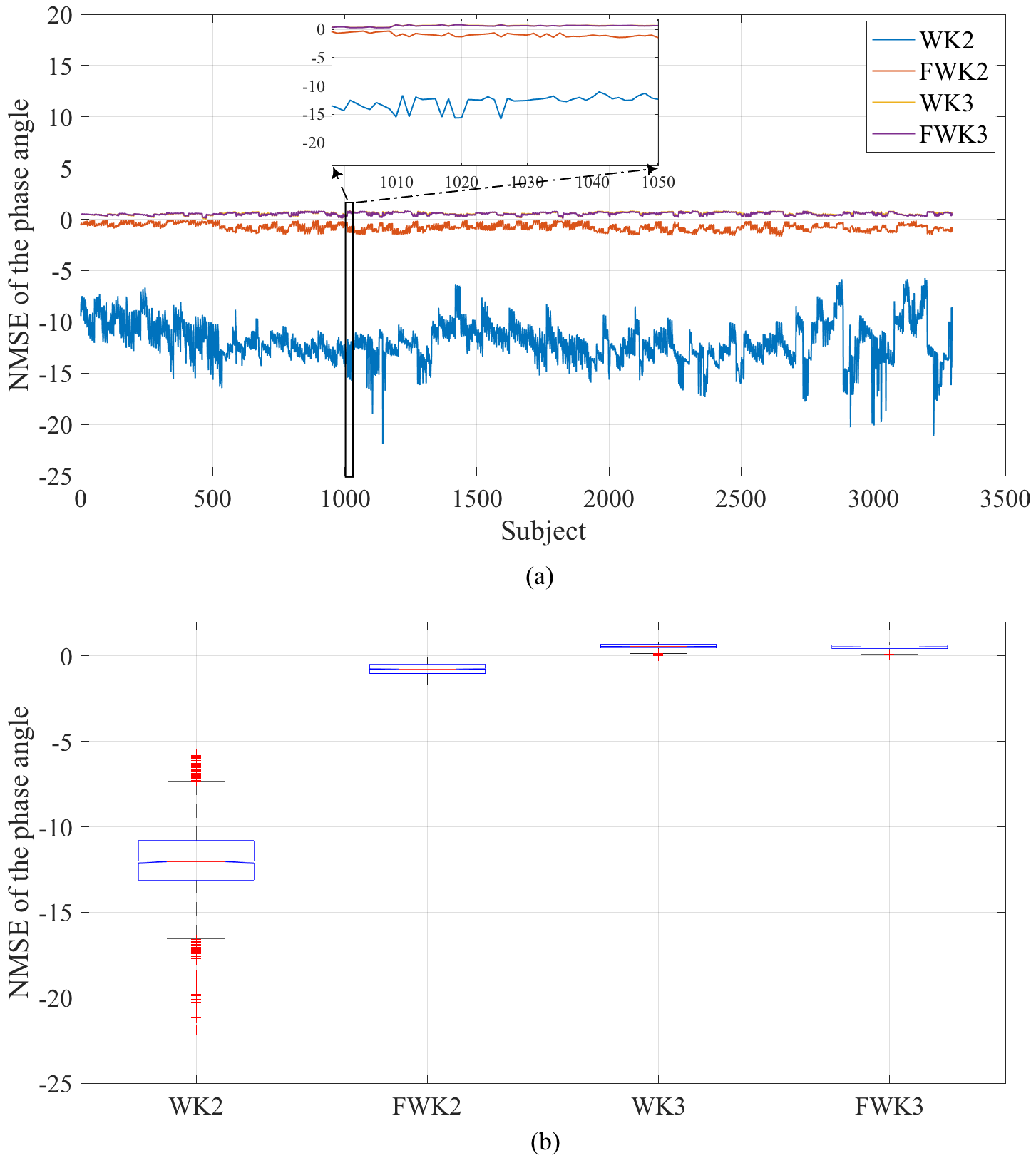}%
	\caption{(a) Comparison of goodness of fit of the aortic input impedance phase angle quantified as $\mathrm {NMSE}$, evaluated for both proposed models (FWK2 and FWK3) and standard Windkessel (WK2 and WK3) for all virtual subject, and (b) box plots providing a visualization of summary statistics of this comparison. }
	\label{fig3}
	\vspace{2mm}
\end{figure}
Fig. 7 (a and b), Fig. 8 (a and b) and Fig. 9 (a and b) show the various comparisons between the performances of the different models. The proposed models $\mathrm {(FWK2\ and \ FWK3)}$ and the standard Windkessel $\mathrm {(WK2\ and\ WK3)}$ are quantified respectively as: 1) $ {REMSE}$  (28) for the overall models, 2) $ {Deviation [\%]}$ (30) to evaluate the deviation of the modulus-based model from the in-silico aortic input impedance modulus, and 3) the $ {NMSE}$ (31) to quantify the error of the phase angle fit. Figs. 7, 8, 9 and table 1. clearly show that, for all subjects, $\mathrm {WK2}$ has the highest $\mathrm {RMSE}$ and $\mathrm {Deviation [\%]}$. Moreover, its $ {(NMSE)}$ is very low, approaching to ($ {-\infty}$). Its mean value is equal to $-11.93 \pm 0.132$, which indicates a poor fit of the phase angle. These findings are in agreement with the results reported in the literature. Indeed, it is known that modulus of the  WK2 decreases to a  small  value and its phase angle converges  to  $\mathrm {-90^{\circ}}$, at medium and high frequencies. However, this convergence property does not corresponds to the observed one using   \textit{in-vivo} measurements, where the modulus decreases  to a plateau value and the phase angle converges  to $\mathrm {0^{\circ}}$. By substituting the ideal capacitor with $\mathrm {FoC}$, in $\mathrm {FWK2}$, we found the fit of the aortic input impedance to be sharply improved; this is illustrated by the decrease of  the mean $\mathrm {RMSE}$ by almost one-half, from $\mathrm {WK2}$ to $\mathrm {FWK2}$, with the $ {Deviation}$ reaching $16.43 \pm 0.33$  vice $61.86 \pm 0.38$ for $\mathrm {WK2}$. Although the $ {NMSE}$ of $\mathrm {(FWK2)}$is not close to one (i.e, good fit), the phase angle pattern is improved, compared to $\mathrm {WK2}$. modulus fitting, $\mathrm {FWK2}$ model appears almost as performant  as $\mathrm {WK3}$, with a $ {Deviation}$ equal to $16.17 \pm 0.31$.
 The proposed $\mathrm {FWK3}$ model provides the best fit, but overall comparable to $\mathrm {WK3}$. All the results presented here, confirm the fact that both $\mathrm {WK3}$ and $\mathrm {FWK2}$ overcome the limitations of the $\mathrm {WK2}$ model, in describing the real input impedance. However, the use of fractional-order element grants a proper measure for the better physiological analysis of the arterial function. In fact, an ideal capacitor can be viewed as a pure storage element that can only imitate a pure elastic behavior, not a viscoelastic one. On the other hand, a fractional-order capacitor lumps both resistive and capacitive properties, in one element, allowing for a reduced-order description of the arterial viscoelastic characteristics. The proposed fractional-order model smoothly incorporates the complex effects and multi-scale properties of vascular tissues, using a reduced-order configuration.
To establish a fair comparison between the estimated compliance of the proposed fractional-order model and its corresponding standard Windkessel model, for both proposed $\mathrm {FWK2}$ and $\mathrm {FWK3}$, we have calculated the effective compliance that can be derived based on equation (15) and from the estimated value of $\tau_{\alpha}$ we calculate $C_\alpha$ as:
\begin{equation}\tag {32}
C_\alpha=\frac{\left(\tau_\alpha\right)^\alpha}{R_p},
\end{equation}
\begin{figure}[!t]
	\centering
	\vspace{0mm}
	\includegraphics[height=10cm, width=12cm]{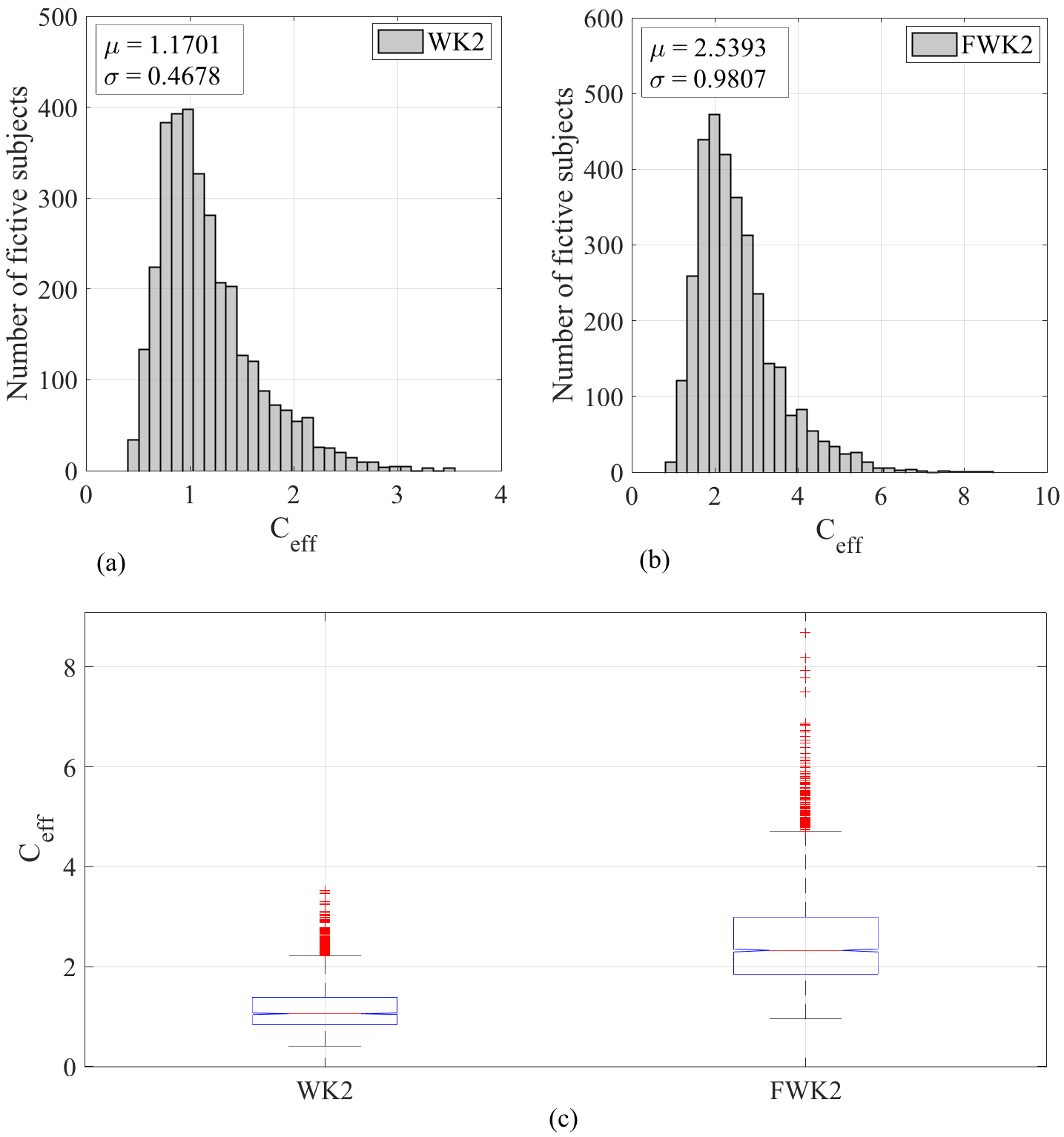}%
	\caption{Distribution of the effective compliance estimates for WK2 in (a) and  FWK2 in (b). (c) box plots, showing a visualization of summary statistics of a comparison between $C_{eff}$ estimates for the two models.}
	\label{fig3}
	\vspace{2mm}
\end{figure}
\begin{figure}[!b]
	\centering
	\vspace{1mm}
	\includegraphics[height=10cm, width=12cm]{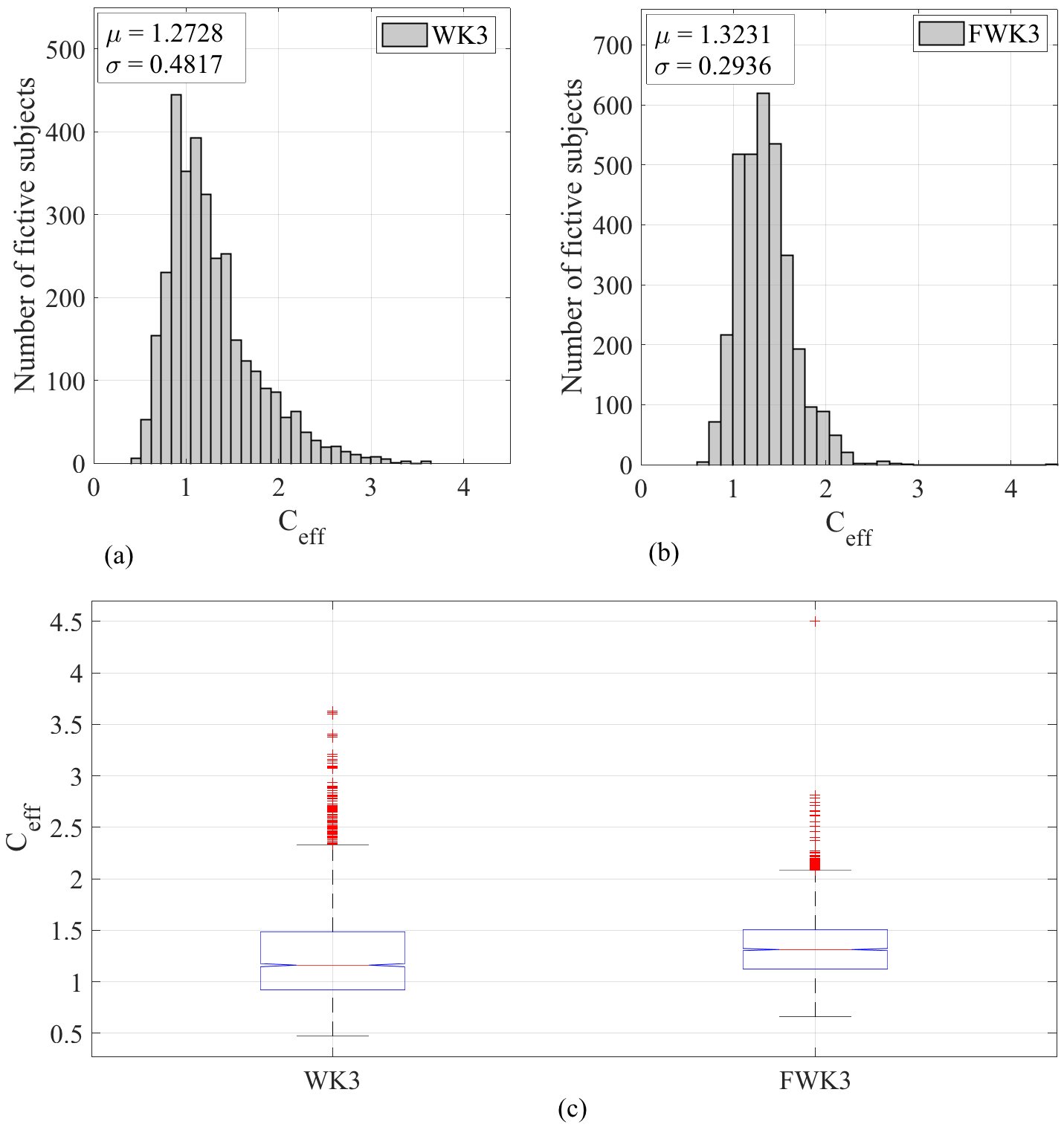}%
	\caption{Distribution of the effective compliance estimates for WK3 in (a) and  FWK3 in (b). (c) box plots, showing a visualization of summary statistics of a comparison between $C_{eff}$ estimates for the two models.}
	\label{fig3}
	\vspace{2mm}
\end{figure}
\begin{figure}[!t]
	\centering
	\vspace{0mm}
	\includegraphics[height=10cm, width=12cm]{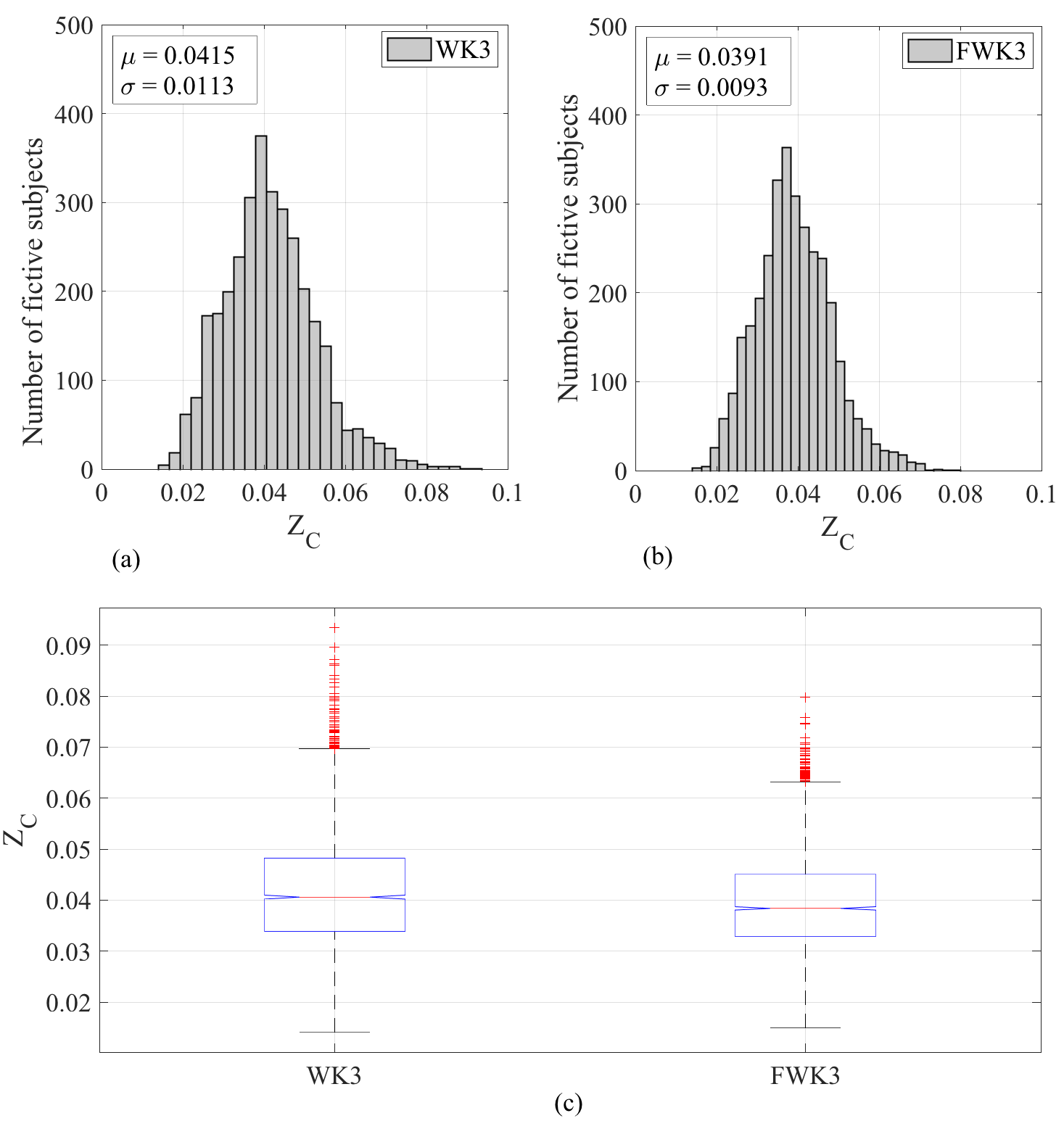}%
	\caption{Distribution of the characteristic impedance estimates for WK3 in (a) and  FWK3 in (b). (c) box plots, showing a visualization of summary statistics of a comparison between $Z_{C}$ estimates for the two models.}
	\label{fig3}
	\vspace{2mm}
\end{figure}
and subsequently extract the effective capacitance representing the effective compliance, is extracted based on (33):
\begin{equation}\tag{33}
C_{eff}=C_\alpha sin \left(\alpha \frac{\pi}{2}\right).
\end{equation}
It is worth noting that, by substituting $\alpha $ by one in (33), $C_{eff}$ will represent the ideal capacitance that corresponds to the standard Windkessel compliance.
The distribution of the estimated effective compliance for $\mathrm {(WK2\ and\ FWK2)}$ are shown in Fig. 10 (a and b); WK3 and FWK3 are shown in Fig. 11 (a and b). It is clear from Fig. 10 (c), that $ {C_{eff}}$ of $\mathrm {FWK2}$ is larger than ${C_{eff}}$ of $\mathrm {WK2}$; however, in Fig. 11 (c), $ {C_{eff}}$ of $\mathrm{FWK3}$ is close to ${C_{eff}}$ of $\mathrm{WK3}$. This result may be explained by the fact that, in the case of FWK2, the value of $\alpha$ is lower than $\mathrm {0.5}$, meaning that the fractional-order element is more resistive than capacitive. Hence, an increase in the effective compliance might be viewed as a compensation of the decrease of the capacitive part, introduced by $\mathrm {FoC}$. However, in the case of $\mathrm {FWK3}$, $ {\alpha}$ is close to $\mathrm {1}$ which induces a reasonable  $ {C_{eff}}$ between WK3 and FWK3.
\begin{table*}[b]
	\vspace{3mm}
	\centering
	\caption{Mean goodness of fit parameters: RMSE, Deviation [\%] and NMSE and mean estimates of $\alpha$, $Z_C$, $C_{eff}$ and $R_p$}
	\renewcommand{\arraystretch}{3}
	\newcolumntype{P}[1]{>{\centering\arraybackslash}p{#1}}
	\resizebox{\linewidth}{!}{	
		\renewcommand{\arraystretch}{2}
		\begin{tabular}{l|l|l|l|l|l|l|l|}
			\cline{2-8}
			&  \multicolumn{3}{l|}{  {\ \ \ \ \ \ {Goodness \ of \ Fit \ Quantification \ Parameters}} }                & \multicolumn{4}{l|}{{ \ \ \ \ \ \ \ \ \ \ \ \  \ \ \ \ \ \ \  \ \ \ Estimated Arterial Model Parameter}}                                          \\ \hline 
			\multicolumn{1}{|l|}{ {Model}}  &{\ \ \ \ \ \ \ RMSE}                 &  {\ \ \ Deviation {[}\%{]} }  &  { \ \ \ \ NMSE   }             & $\mathrm{\ \ \ \ \ \  \ \alpha}$ & $\mathrm{\ \ \ \ \ \ \ \ \ \ \ \ Z_C}$                   & \textbf {$\mathrm{\ \ \ \ \ \ C_{eff}}$}          & \textbf {$\mathrm{\ \ \ \ \ \ R_P}$ }                 \\ \hline \hline 
			\multicolumn{1}{|l|}{\textbf {WK2}}  & 0.0455 $\pm 8.35e\!\!-\!\!4$ & 61.86 $\pm 0.3898$ & -11.93 $\pm 0.132$ & \cellcolor{gray!20}                      &  \cellcolor{gray!20}                      & 1.1701 $\pm 0.03$ & 0.9733 $\pm 0.0051$ \\ \hline 
			\multicolumn{1}{|l|}{\textbf {FWK2}} & 0.0263 $\pm 3.93e\!\!-\!\!4$ & 16.43 $\pm 0.3363$ & -0.76 $\pm 0.0243$ & 0.3687$\pm 0.0081$    &  \cellcolor{gray!20}                   & 2.5393 $\pm 0.06$ & 0.9733 $\pm 0.0051$ \\ \hline \hline 
			\multicolumn{1}{|l|}{\textbf {WK3}}  & 0.0176 $\pm 3.76e\!\!-\!\!4$ & 16.17 $\pm 0.3128$ & 0.55 $\pm 0.0099$   &  \cellcolor{gray!20}                     & 0.0415 $\pm 7.74e\!\!-\!\!4$  & 1.2728 $\pm 0.03$ & 0.9733 $\pm 0.0051$ \\ \hline 
			\multicolumn{1}{|l|}{\textbf {FWK3}} & 0.0170 $\pm 3.57e\!\!-\!\!4$ & 15.35 $\pm 0.3310$  & 0.52 $\pm 0.0095$   & 0.9525 $\pm 0.0050$   & 0.03911 $\pm 6.39e\!\!-\!\!4$ & 1.3431 $\pm 0.02$ & 0.9733 $\pm 0.0051$ \\ \hline 
		\end{tabular}
	}
\end{table*}
\begin{figure*}[t!]
	\centering
	\vspace{0mm}
	\includegraphics[height=18cm,width=18cm]{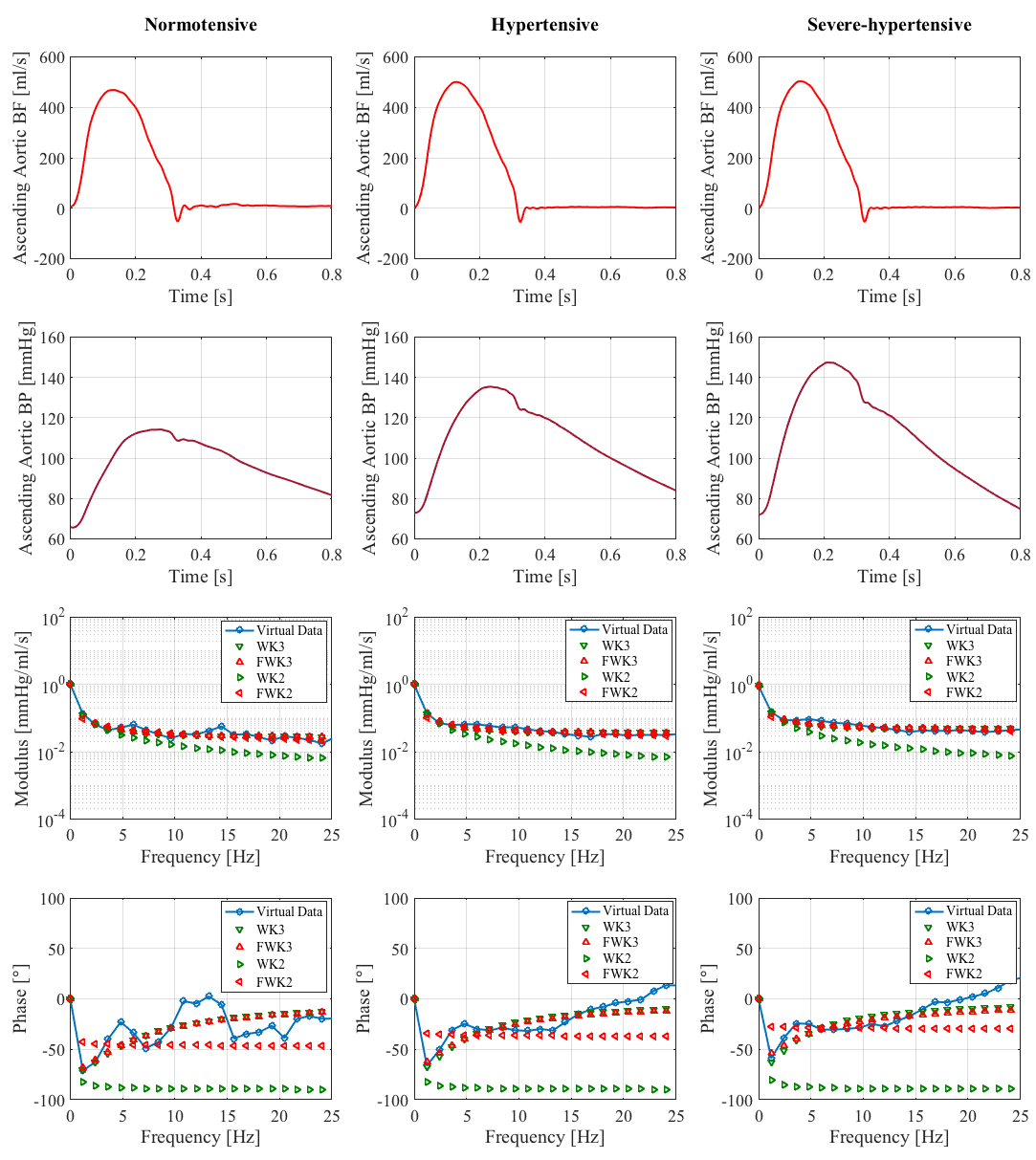}%
	\caption{The aortic input impedance validation using in-silico human data for different physiological state (normotensive, hypertensive and severe-hypertensive). The figure shows the blood pressure (BP) and flow (BF) at the level of the ascending aorta in time domain and the corresponding aortic input impedance modulus (presented in log-scale) and phase angle as a function of frequency. It also shows a comparison between the reconstructed impedance modulus and phase angle, based on WK2 and FWK2 models.}
	\label{fig3}
	\vspace{2mm}
\end{figure*}
Fig. 12, shows a comparative illustration of the estimated characteristic impedance $ {Z_C}$ for both $\mathrm {WK3}$ and  $\mathrm {FWK3}$ models. It is clear from these results that, in both cases, we have relatively equal estimated values. In general, the effect of adding $ {Z_C}$ to the $\mathrm {FWK2}$ configuration can be mostly visualized in the correction of the phase angle pattern; however, in the case of $\mathrm {WK2}$, it affects both the phase angle and modulus. Future, a deep investigation of the relationship between $\mathrm {FoC}$ and the characteristic impedance will be conducted.
\begin{figure}[!t]
	\centering
	\vspace{0mm}
	\includegraphics[height=11cm, width=13cm]{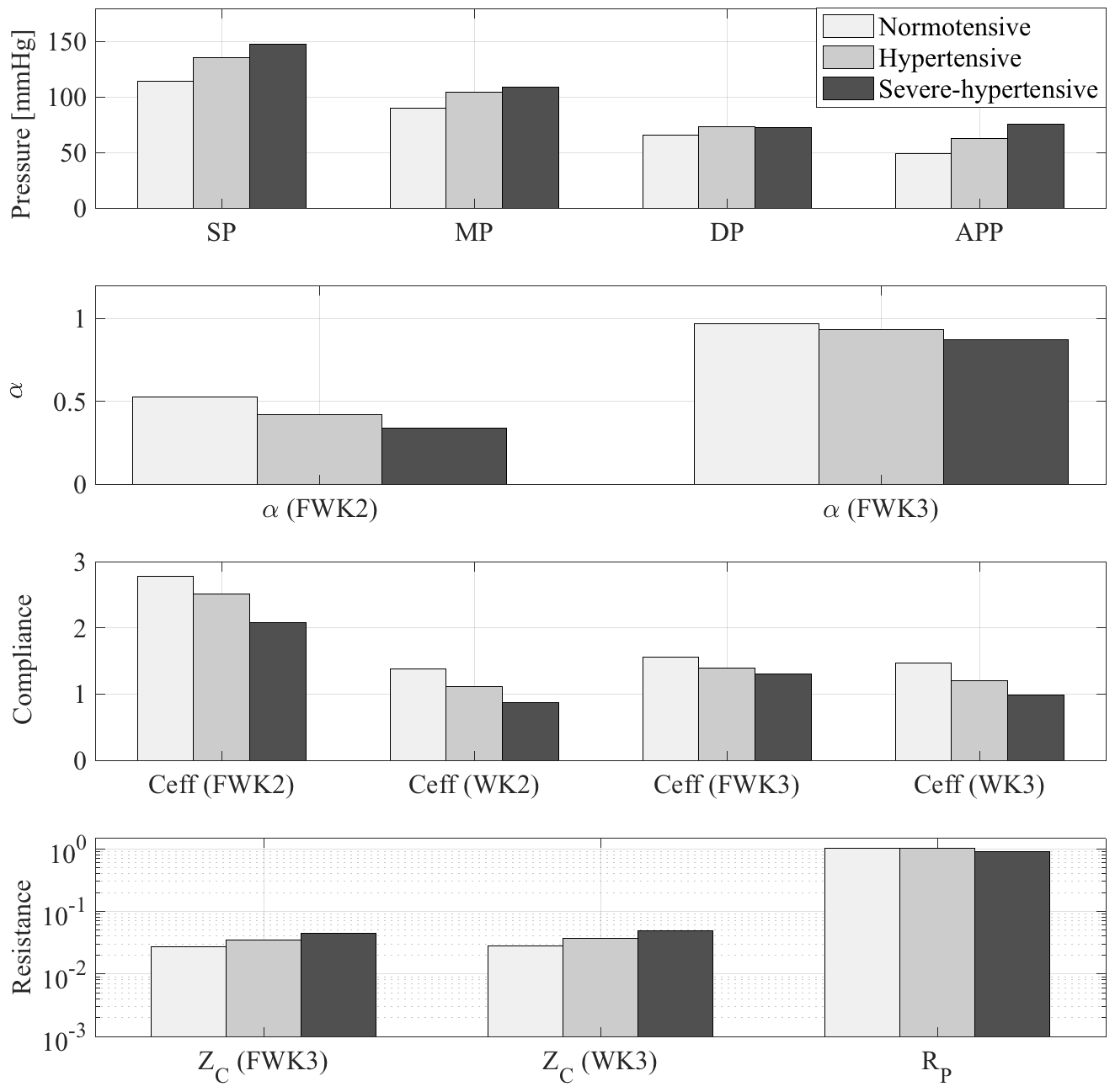}%
	\caption{ Bar graph of the arterial parameter (SBP: systolic aortic blood pressure, MP: mean aortic blood pressure, DBP: Diastolic aortic blood pressure  and APP: aortic pulse pressure) and the estimated standard and fractional-order proposed models for three different physiological states (Normotensive, hypertensive and sever-hypertensive).}
	\label{fig3}
	\vspace{4mm}
\end{figure}
\begin{table}[b]
	\vspace{3mm}
	\caption{ Diastolic, systolic, pulse blood pressure and the corresponding estimated value of the fractional differentiation order parameter \text {$\alpha$} of 3 different physiological states.}
	\centering
	\resizebox{1\linewidth}{!}{
		\renewcommand{\arraystretch}{1.3}
		\begin{tabular}{cccccc}
			\toprule
			& Diastolic Pressure [mmHg] & Systolic Pressure [mmHg] &Pulse Pressure [mmHg]& $ {\alpha}$ (FWK2)& $\mathrm {\alpha}$ (FWK3) \\ \midrule
			Normotensive       & 65.63             & 114.23            & 48.60          & 0.52 & 0.96    \\\midrule
			Hypertensive       & 72.79             & 135.45            & 62.46         & 0.42 & 0.93    \\ \midrule
			Severe-Hypertensive & 72.16             & 147.17            & 75.24          & 0.33 & 0.87      \\ \bottomrule
		\end{tabular}
	}
	\vspace{4mm}
\end{table}
Fig. 13 shows, an example of in-silico aortic input impedance patterns, for different physiological states (normotensive, hypertensive and sever-hypertensive) with their corresponding aortic blood pressure and flow waveforms. The diastolic, systolic, pulse pressure values and the estimated $\mathrm {\alpha}$ parameters, for each state, are reported in Table.2. Additionally, Fig. 14 represents the bar graphs of the hemodynamic parameters, as well as the estimated parameters of both the proposed models and the standard Windkessel. These results are in consistent with the conclusions aforementioned. In fact, it shows that $\mathrm {FWK2}$ model is significantly improving the prediction of the in-silico input impedance modulus, performing closely to $\mathrm {WK3}$ and $\mathrm {FWK3}$. The  variation of $\mathrm{\alpha}$ values for both $\mathrm {FWK2}$ and $\mathrm {FWK3}$, from a physiological state to another, show that the differentiation order ${\alpha}$ of the fractional-order operator is correlated with all the arterial parameter. For instance, the (${\alpha}$) values, systolic blood pressure ($\mathrm{SBP}$) or aortic pulse pressure ($\mathrm{APP}$) show a negative correlation , and as a consequence, an increase of SBP, from one normotensive state to a severe-hypertensive, $\mathrm{SBP}$ increases. However ${\alpha}$ decreases for both $\mathrm{FWK3}$ and $\mathrm{FWK3}$. Hence, the fractional differentiation order may implicate a physiological insight. In fact, a decrease of $\mathrm{\alpha}$ means an increase of the resistive part. On the other hand, physiologically, one of the acute causes of high blood pressure is arterial stiffness \cite{wang2016viscoelastic}. Accordingly, the new parameter $\alpha$ can be investigated as a bio-marker that can lump the overall viscoelasticity properties of the human arterial tree. It may also help to better understand the arterial stiffness dependencies.
\section{Conclusion}
In this study, we have conducted an investigation of the arterial Windkessel within a fractional-order modeling framework. Tools from fractional-order calculus such as the fractional-order impedance have been used to estimate and measure the aortic input impedance. In this paper, we introduce two fractional-order model analog circuits that incorporate a fractional-order capacitor to assess the arterial input impedance, at the level of the ascending aorta. The fractional-order element lumps both resistive and capacitive properties that represent the viscoelastic behavior of the systemic arteries. The contribution of both characteristics can be controlled by the fractional differentiation order, $\alpha$, enabling an accurate, reduced-order and flexible description.
The validation and comparison with the conventional $\mathrm{WK}$ results show that the proposed models provide a significant  improvement of the standard Windkessel. It also reveals that the fractional differentiation operator order $\alpha$ may have a powerful role, as a physiological bio-marker.

In the future, we aim to prove that such a simplified fractional-order model can smoothly incorporate the complex effects and the multi-scale properties of the cardiovascular bio-tissues, using reduced-order configurations. This new paradigm may open new avenues towards a better understanding of the arterial stiffness dependencies, particularly and introduce for a new hemodynamic problems resolving framework, generally. 
\section*{Acknowledgment}
The research reported in this publication was supported by King Abdullah University of Science and Technology (KAUST) Base Research Fund, (BAS/1/1627-01-01). Additionally, the authors would like to thank Dr. Ali Haneef, associate consultant cardiac surgeon and co-chairman quality management at King Faisal Cardiac Center, King Abdulaziz Medical City, National Guard Health Affairs, in the Western Region, Jeddah, KSA for his assistance and valuable advices.

\bibliographystyle{apa}
\bibliography{bibfile_AJP}

\begin{thebibliography}{}

\bibitem[\protect\astroncite{Aboelkassem and
  Virag}{2019}]{aboelkassem2019hybrid}
Aboelkassem, Y. and Virag, Z. (2019).
\newblock A hybrid windkessel-womersley model for blood flow in arteries.
\newblock {\em Journal of theoretical biology}, 462:499--513.

\bibitem[\protect\astroncite{Bahloul and Laleg-Kirati}{2018}]{bahloul2018three}
Bahloul, M.~A. and Laleg-Kirati, T.~M. (2018).
\newblock Three-element fractional-order viscoelastic arterial windkessel
  model.
\newblock In {\em 2018 40th Annual International Conference of the IEEE
  Engineering in Medicine and Biology Society (EMBC)}, pages 5261--5266. IEEE.

\bibitem[\protect\astroncite{Bahloul and Laleg-Kirati}{2019}]{bahloul2019two}
Bahloul, M.~A. and Laleg-Kirati, T.~M. (2019).
\newblock Two-element fractional-order windkessel model to assess the arterial
  input impedance.
\newblock In {\em accepted in 2019 41th Annual International Conference of the
  IEEE Engineering in Medicine and Biology Society (EMBC)}. IEEE.

\bibitem[\protect\astroncite{Balocco et~al.}{2010}]{balocco2010estimation}
Balocco, S., Basset, O., Courbebaisse, G., Boni, E., Frangi, A.~F., Tortoli,
  P., and Cachard, C. (2010).
\newblock Estimation of the viscoelastic properties of vessel walls using a
  computational model and doppler ultrasound.
\newblock {\em Physics in Medicine \& Biology}, 55(12):3557.

\bibitem[\protect\astroncite{Borlaug and Kass}{2011}]{borlaug2011ventricular}
Borlaug, B.~A. and Kass, D.~A. (2011).
\newblock Ventricular--vascular interaction in heart failure.
\newblock {\em Cardiology clinics}, 29(3):447--459.

\bibitem[\protect\astroncite{Burattini and
  Natalucci}{1998}]{burattini1998complex}
Burattini, R. and Natalucci, S. (1998).
\newblock Complex and frequency-dependent compliance of viscoelastic windkessel
  resolves contradictions in elastic windkessels.
\newblock {\em Medical engineering \& physics}, 20(7):502--514.

\bibitem[\protect\astroncite{Burattini
  et~al.}{1999}]{burattini1999viscoelasticity}
Burattini, R., Natalucci, S., and Campbell, K.~B. (1999).
\newblock Viscoelasticity modulates resonance in the terminal aortic
  circulation.
\newblock {\em Medical engineering \& physics}, 21(3):175--185.

\bibitem[\protect\astroncite{{\v{C}}ani{\'c} et~al.}{2006}]{vcanic2006modeling}
{\v{C}}ani{\'c}, S., Tamba{\v{c}}a, J., Guidoboni, G., Mikeli{\'c}, A.,
  Hartley, C.~J., and Rosenstrauch, D. (2006).
\newblock Modeling viscoelastic behavior of arterial walls and their
  interaction with pulsatile blood flow.
\newblock {\em SIAM Journal on Applied Mathematics}, 67(1):164--193.

\bibitem[\protect\astroncite{Capoccia}{2015}]{capoccia2015development}
Capoccia, M. (2015).
\newblock Development and characterization of the arterial w indkessel and its
  role during left ventricular assist device assistance.
\newblock {\em Artificial organs}, 39(8):E138--E153.

\bibitem[\protect\astroncite{Craiem and Armentano}{2007}]{ref8}
Craiem, D. and Armentano, R.~L. (2007).
\newblock A fractional derivative model to describe arterial viscoelasticity.
\newblock {\em Biorheology}, 44(4):251--263.

\bibitem[\protect\astroncite{Craiem et~al.}{2008a}]{ref9}
Craiem, D., Rojo, F., Atienza, J., Guinea, G., and Armentano, R.~L. (2008a).
\newblock Fractional calculus applied to model arterial viscoelasticity.
\newblock {\em Latin American applied research}, 38(2):141--145.

\bibitem[\protect\astroncite{Craiem et~al.}{2008b}]{ref10}
Craiem, D., Rojo, F.~J., Atienza, J.~M., Armentano, R.~L., and Guinea, G.~V.
  (2008b).
\newblock Fractional-order viscoelasticity applied to describe uniaxial stress
  relaxation of human arteries.
\newblock {\em Physics in medicine and biology}, 53(17):4543.

\bibitem[\protect\astroncite{Doehring et~al.}{2005}]{ref7}
Doehring, T.~C., Freed, A.~D., Carew, E.~O., and Vesely, I. (2005).
\newblock Fractional order viscoelasticity of the aortic valve cusp: an
  alternative to quasilinear viscoelasticity.
\newblock {\em Journal of biomechanical engineering}, 127(4):700--708.

\bibitem[\protect\astroncite{Elwakil}{2010}]{ref21}
Elwakil, A.~S. (2010).
\newblock Fractional-order circuits and systems: An emerging interdisciplinary
  research area.
\newblock {\em IEEE Circuits and Systems Magazine}, 10(4):40--50.

\bibitem[\protect\astroncite{Freeborn}{2013}]{freeborn2013survey}
Freeborn, T.~J. (2013).
\newblock A survey of fractional-order circuit models for biology and
  biomedicine.
\newblock {\em IEEE Journal on emerging and selected topics in circuits and
  systems}, 3(3):416--424.

\bibitem[\protect\astroncite{Hemmer et~al.}{2009}]{hemmer2009role}
Hemmer, J.~D., Nagatomi, J., Wood, S.~T., Vertegel, A.~A., Dean, D., and
  LaBerge, M. (2009).
\newblock Role of cytoskeletal components in stress-relaxation behavior of
  adherent vascular smooth muscle cells.
\newblock {\em Journal of biomechanical engineering}, 131(4):041001.

\bibitem[\protect\astroncite{Hollkamp et~al.}{2018}]{hollkamp2018model}
Hollkamp, J.~P., Sen, M., and Semperlotti, F. (2018).
\newblock Model-order reduction of lumped parameter systems via fractional
  calculus.
\newblock {\em Journal of Sound and Vibration}, 419:526--543.

\bibitem[\protect\astroncite{Holzapfel et~al.}{2002}]{holzapfel2002structural}
Holzapfel, G.~A., Gasser, T.~C., and Stadler, M. (2002).
\newblock A structural model for the viscoelastic behavior of arterial walls:
  continuum formulation and finite element analysis.
\newblock {\em European Journal of Mechanics-A/Solids}, 21(3):441--463.

\bibitem[\protect\astroncite{Ionescu et~al.}{2011}]{ref22}
Ionescu, C.~M., Machado, J.~T., and De~Keyser, R. (2011).
\newblock Modeling of the lung impedance using a fractional-order ladder
  network with constant phase elements.
\newblock {\em IEEE Transactions on biomedical circuits and systems},
  5(1):83--89.

\bibitem[\protect\astroncite{Jaishankar and
  McKinley}{2013}]{jaishankar2013power}
Jaishankar, A. and McKinley, G.~H. (2013).
\newblock Power-law rheology in the bulk and at the interface: quasi-properties
  and fractional constitutive equations.
\newblock {\em Proceedings of the Royal Society A: Mathematical, Physical and
  Engineering Sciences}, 469(2149):20120284.

\bibitem[\protect\astroncite{Jumarie}{2006}]{jumarie2006modified}
Jumarie, G. (2006).
\newblock Modified riemann-liouville derivative and fractional taylor series of
  nondifferentiable functions further results.
\newblock {\em Computers \& Mathematics with Applications},
  51(9-10):1367--1376.

\bibitem[\protect\astroncite{Kilbas et~al.}{2006}]{kilbas2006theory}
Kilbas, A. A.~A., Srivastava, H.~M., and Trujillo, J.~J. (2006).
\newblock {\em Theory and applications of fractional differential equations},
  volume 204.
\newblock Elsevier Science Limited.

\bibitem[\protect\astroncite{Kobayashi et~al.}{2012}]{kobayashi2012modeling}
Kobayashi, Y., Kato, A., Watanabe, H., Hoshi, T., Kawamura, K., and Fujie,
  M.~G. (2012).
\newblock Modeling of viscoelastic and nonlinear material properties of liver
  tissue using fractional calculations.
\newblock {\em Journal of Biomechanical Science and Engineering},
  7(2):177--187.

\bibitem[\protect\astroncite{Machado}{2001}]{ref29}
Machado, J. (2001).
\newblock Discrete-time fractional-order controllers.
\newblock {\em Fractional Calculus and Applied Analysis}, 4:47--66.

\bibitem[\protect\astroncite{Magin}{2006}]{magin2006fractional}
Magin, R.~L. (2006).
\newblock {\em Fractional calculus in bioengineering}.
\newblock Begell House Redding.

\bibitem[\protect\astroncite{Milnor}{1989}]{milnor1989vascular}
Milnor, W. (1989).
\newblock Vascular impedance.
\newblock {\em Hemodynamics, 2nd Ed., Baltimore, MD, Williams \& Wilkins},
  pages 167--203.

\bibitem[\protect\astroncite{Milnor}{1975}]{milnor1975arterial}
Milnor, W.~R. (1975).
\newblock Arterial impedance as ventricular afterload.
\newblock {\em Circulation Research}, 36(5):565--570.

\bibitem[\protect\astroncite{Naghibolhosseini and
  Long}{2018}]{naghibolhosseini2018fractional}
Naghibolhosseini, M. and Long, G.~R. (2018).
\newblock Fractional-order modelling and simulation of human ear.
\newblock {\em International Journal of Computer Mathematics},
  95(6-7):1257--1273.

\bibitem[\protect\astroncite{Noordergraaf}{2012}]{noordergraaf2012circulatory}
Noordergraaf, A. (2012).
\newblock {\em Circulatory system dynamics}, volume~1.
\newblock Elsevier.

\bibitem[\protect\astroncite{O'Rourke et~al.}{1984}]{o1984physiological}
O'Rourke, M.~F., Yaginuma, T., and Avolio, A.~P. (1984).
\newblock Physiological and pathophysiological implications of
  ventricular/vascular coupling.
\newblock {\em Annals of biomedical engineering}, 12(2):119--134.

\bibitem[\protect\astroncite{Podlubny}{1998}]{podlubny1998fractional}
Podlubny, I. (1998).
\newblock {\em Fractional differential equations: an introduction to fractional
  derivatives, fractional differential equations, to methods of their solution
  and some of their applications}, volume 198.
\newblock Elsevier.

\bibitem[\protect\astroncite{Quick et~al.}{2001a}]{quick2001constructive}
Quick, C.~M., Berger, D.~S., and Noordergraaf, A. (2001a).
\newblock Constructive and destructive addition of forward and reflected
  arterial pulse waves.
\newblock {\em American Journal of Physiology-Heart and Circulatory
  Physiology}, 280(4):H1519--H1527.

\bibitem[\protect\astroncite{Quick et~al.}{2006}]{quick2006resolving}
Quick, C.~M., Berger, D.~S., Stewart, R.~H., Laine, G.~A., Hartley, C.~J., and
  Noordergraaf, A. (2006).
\newblock Resolving the hemodynamic inverse problem.
\newblock {\em IEEE transactions on biomedical engineering}, 53(3):361--368.

\bibitem[\protect\astroncite{Quick et~al.}{2001b}]{quick2001infinite}
Quick, C.~M., Young, W.~L., and Noordergraaf, A. (2001b).
\newblock Infinite number of solutions to the hemodynamic inverse problem.
\newblock {\em American Journal of Physiology-Heart and Circulatory
  Physiology}, 280(4):H1472--H1479.

\bibitem[\protect\astroncite{Reesink and
  Spronck}{2018}]{reesink2018constitutive}
Reesink, K.~D. and Spronck, B. (2018).
\newblock Constitutive interpretation of arterial stiffness in clinical
  studies: a methodological review.
\newblock {\em American Journal of Physiology-Heart and Circulatory
  Physiology}, 316(3):H693--H709.

\bibitem[\protect\astroncite{Sharp et~al.}{2000}]{sharp2000aortic}
Sharp, M.~K., Pantalos, G.~M., Minich, L., Tani, L.~Y., McGough, E.~C., and
  Hawkins, J.~A. (2000).
\newblock Aortic input impedance in infants and children.
\newblock {\em Journal of applied physiology}, 88(6):2227--2239.

\bibitem[\protect\astroncite{Shi et~al.}{2011}]{shi2011review}
Shi, Y., Lawford, P., and Hose, R. (2011).
\newblock Review of zero-d and 1-d models of blood flow in the cardiovascular
  system.
\newblock {\em Biomedical engineering online}, 10(1):33.

\bibitem[\protect\astroncite{Stergiopulos
  et~al.}{1995}]{stergiopulos1995evaluation}
Stergiopulos, N., Meister, J., and Westerhof, N. (1995).
\newblock Evaluation of methods for estimation of total arterial compliance.
\newblock {\em American Journal of Physiology-Heart and Circulatory
  Physiology}, 268(4):H1540--H1548.

\bibitem[\protect\astroncite{Vlachopoulos
  et~al.}{2011}]{vlachopoulos2011mcdonald}
Vlachopoulos, C., O'Rourke, M., and Nichols, W.~W. (2011).
\newblock {\em McDonald's blood flow in arteries: theoretical, experimental and
  clinical principles}.
\newblock CRC press.

\bibitem[\protect\astroncite{Wang et~al.}{2016}]{wang2016viscoelastic}
Wang, Z., Golob, M.~J., and Chesler, N.~C. (2016).
\newblock Viscoelastic properties of cardiovascular tissues.
\newblock In {\em Viscoelastic and viscoplastic materials}. IntechOpen.

\bibitem[\protect\astroncite{Westerhof et~al.}{2009}]{westerhof2009arterial}
Westerhof, N., Lankhaar, J.-W., and Westerhof, B.~E. (2009).
\newblock The arterial windkessel.
\newblock {\em Medical \& biological engineering \& computing}, 47(2):131--141.

\bibitem[\protect\astroncite{Westerhof et~al.}{2019}]{westerhof2019arterial}
Westerhof, N., Stergiopulos, N., Noble, M.~I., and Westerhof, B.~E. (2019).
\newblock Arterial input impedance.
\newblock In {\em Snapshots of Hemodynamics}, pages 195--206. Springer.

\bibitem[\protect\astroncite{Willemet et~al.}{2015}]{willemet2015database}
Willemet, M., Chowienczyk, P., and Alastruey, J. (2015).
\newblock A database of virtual healthy subjects to assess the accuracy of
  foot-to-foot pulse wave velocities for estimation of aortic stiffness.
\newblock {\em American Journal of Physiology-Heart and Circulatory
  Physiology}, 309(4):H663--H675.

\bibitem[\protect\astroncite{Xiao et~al.}{2017}]{xiao2017arterial}
Xiao, H., Tan, I., Butlin, M., Li, D., and Avolio, A.~P. (2017).
\newblock Arterial viscoelasticity: role in the dependency of pulse wave
  velocity on heart rate in conduit arteries.
\newblock {\em American Journal of Physiology-Heart and Circulatory
  Physiology}.

\end{thebibliography}
\vspace{-12mm}
\end{document}